\documentclass[prb,aps,superscriptaddress,twocolumn,longbibliography]{revtex4-2}
\usepackage{graphicx}
\usepackage{bm}
\usepackage{braket}
\usepackage{amsmath}
\usepackage{amsfonts}
\usepackage{amssymb}
\usepackage{epsfig}
\usepackage{color}
\usepackage[colorlinks,linkcolor=blue,citecolor=blue]{hyperref}
\usepackage{ulem}
\usepackage{lineno}
\usepackage{comment}
\begin{document}

\title{Spin and orbital Edelstein effect in spin-orbit coupled noncentrosymmetric superconductors}

\author{Satoshi Ando}
\affiliation{Dipartimento di Fisica ``E. R. Caianiello", Universit\`a di Salerno, IT-84084 Fisciano (SA), Italy}
\affiliation{Department of Applied Physics, Nagoya University, Nagoya 464-8603, Japan}

\author{Yukio Tanaka}
\affiliation{Department of Applied Physics, Nagoya University, Nagoya 464-8603, Japan}

\author{Mario Cuoco}
\affiliation{CNR-SPIN, I-84084 Fisciano (Salerno), Italy, c/o Universit\'a di Salerno, I-84084 Fisciano (Salerno), Italy}

\author{Luca Chirolli}
\affiliation{NEST, Istituto Nanoscienze-CNR and Scuola Normale Superiore, I-56127 Pisa, Italy}
\affiliation{Quantum Research Center, Technology Innovation Institute, P.O. Box 9639 Abu Dhabi (UAE)}

\author{Maria Teresa Mercaldo}
\affiliation{Dipartimento di Fisica ``E. R. Caianiello", Universit\`a di Salerno, IT-84084 Fisciano (SA), Italy}

\begin{abstract}
Superconductors without inversion symmetry can exhibit a non-zero magnetization when a supercurrent is present, leading to non-dissipative magnetoelectric effects. Here, we focus on understanding the relation between the spin and orbital properties of these effects in spin-singlet noncentrosymmetric superconductors with orbital Rashba coupling. Our findings indicate that the correlation between spin and orbital moments induced by supercurrents is generally not determined by the sign of atomic spin-orbit coupling, but rather by the number of bands at the Fermi level and the orbital characteristics of those bands with respect to the mirror parity. 
The outcomes demonstrate that the sign change of the orbital Edelstein effect near avoiding crossing bands remains robust despite modifications to the atomic spin-orbit amplitude. Furthermore, we observe that the spin Edelstein effect is typically one order of magnitude smaller than the orbital Edelstein effect, but can be significantly enhanced in certain scenarios with increased atomic spin-orbit coupling.  
\end{abstract}

\maketitle

\section{Introduction}

Spintronics and orbitronics are emerging fields of research that involve the control and manipulation of the spin and orbital angular momentum (OAM) of electrons in materials and aim to advance the knowledge of spin and OAM dynamics as well as their mutual coupling.
The OAM is a crucial characteristic of electrons in atoms and serves as a fundamental degree of freedom in the framework of orbitronics \cite{Go_2021,Jo2024, Burgos2024,Phong2019,Tokatly2010}. In solids, indeed, the OAM has been traditionally thought to be suppressed by the crystalline field potential and can only weakly manifest through spin-orbit coupling (SOC) in materials with spin splitting. 
However, recent observations \cite{Choi2023,Kontani2009,Go2018,Sunko2017,Park2011} have called into question the idea that spin-orbit coupling is the sole contributor to the generation of orbital angular momentum. They have instead shown that OAM can exist even without SOC, and that the strength of OAM responses can surpass that of spin responses.
Moreover, the spin characteristics in materials with sizable spin-orbit coupling could directly originate from orbital angular momentum properties \cite{Go2018,Sunko2017,Park2011}.
When combined with superconductivity, achievements in these fields can have the potential to revolutionize the way one can store and process information by exploiting the intrinsic quantum coherent properties of the Cooper paired state and its ability to conduct electricity without resistance when cooled to low temperatures. By incorporating orbitronic \cite{AQT2023} and spin-orbit effects into superconducting materials or hybrids \cite{Amundsen2024}, for instance, a design of novel types of electronic devices with increased speed and energy efficiency can be envisioned. 
Orbital degrees of freedom are particularly relevant in acentric superconductors due to the presence of the so-called 
orbital Rashba coupling~\cite{Park2011,park12,Khalsa2013PRB,kim14,Mercaldo2020}. 
In systems that are two-dimensional (2D) or quasi-two-dimensional, the lack of inversion symmetry results in the absence of horizontal mirror symmetry $\mathcal{M}_z$, setting out 
a polar axis ${\hat z}$ and leading to an orbital Rashba coupling $\alpha$ that mixes orbital configurations with different $\mathcal{M}_z$ mirror parity. Similar to the well-known Rashba spin-orbit coupling, the orbital Rashba term connects the atomic angular momentum ${\bf L}$ with the crystal wave-vector $\bf{k}$ ~\cite{Park2011,park12,Khalsa2013PRB,kim14,Mercaldo2020}.
The orbital Rashba coupling in layered superconductors has been demonstrated to trigger a reconstruction of the superconducting phase. 
This transition results in a $\pi$-paired phase characterized by a modification in the sign of the superconducting order parameter across different orbitals \cite{Mercaldo2020}.
The occurrence of orbital antiphase pairing can have significant implications for tunneling, magnetic response, and Josephson effects \cite{Mercaldo2021_a,Bours2020,Guarcello2022,Yerin2022,Yerin2024}. In 2D spin-singlet superconductors with distinct low crystalline symmetry, stable vortices carrying orbital angular momentum have been predicted \cite{Mercaldo2022} as well as high-orbital moment Cooper pairs \cite{AQT2023}. Regarding spin-triplet pairing of different orbitals, acentric multiorbital superconductors can exhibit an even-parity s-wave topological phase, leading to 0-, $\pi$-, and $\phi$- Josephson phase couplings with high-harmonics in the current phase relation \cite{fukaya18, fukaya19, fukaya20, fukaya22}. 

A particularly remarkable phenomenon occurs when inversion symmetry is broken in systems such as surfaces, interfaces, or crystals without inversion symmetry: the use of an external electric field or charge current leads to the creation of a uniform spin polarization as a result of Rashba spin-orbit coupling \cite{Edelstein1990,AronovLyanda1989}. This phenomenon is commonly referred to as the Edelstein effect \cite{Edelstein1990}, or alternatively as the Aronov–Lyanda-Geller–Edelstein effect \cite{AronovLyanda1989}. In addition to Rashba spin-orbit systems, the Edelstein effect is recently receiving a considerable attention in a large variety of physical platforms and materials\cite{Johansson_2024}. In fact it has been investigated in Weyl semimetals \cite{Zhao2020}, chiral crystals \cite{Vorobev1979,Shalygin2012,Tenzin2023,Roy2022,Furukawa2017}, oxides interface \cite{Vaz2019, Seibold2017,Johansson2021,Trama2022,ElHamdi2023}, topological insulators \cite{Culcer2010,Mellnik2014,Li2014}, and transition metal dichalcogenides \cite{Bentez2020}. 

Differently from normal conductors, in noncentrosymmetric superconductors the supercurrents can result in the generation of a static magnetization without any dissipation \cite{Edelstein1995,Edelstein2005}. 
Then, it has been also predicted the possibility of an inverse Edelstein effect \cite{Yip2002,Dimitrova2003,Samokhin2004,Konschelle2015,Dimitrova2007}.
While much attention has been given to the spin degrees of freedom in spin-Rashba-type superconductors \cite{Edelstein2005,Fujimoto2005,Konschelle2015,Tkachov2017,He2019,Ilic2020}, including all gyrotropic crystal point groups \cite{He2020} and topological superconducting systems \cite{Ikeda2020}, the study has been extended to other noncentrosymmetric superconductors highlighting the contribution of electrons' orbital moments and multiorbital degrees of freedom \cite{He2021,Chirolli2022}. Moroever, the study of the optical response of noncentrosymmetric superconductors indicates a relation with the Edelstein effect \cite{Shinada2023}.
Edelstein effects have been also predicted to play a relevant role in connection with unconventional magnetic states, 
\cite{Zyuzin2024}
in cuprate superconductors \cite{Raines2019}, in superconducting hybrid structures with quantum spin Hall insulators \cite{Bobkova2016,Tkachov2017}, and with triplet Cooper pairs \cite{Amundsen2017}.
Regarding the experimental observations, recent findings have attributed the asymmetric Fraunhofer patterns in Nb based nanojunctions to the Rashba-Edelstein spin density generated at the interface \cite{Senapati2023}. Moreover,
the superconducting nonreciprocal transport observed in planar Josephson junctions of MoTe$_2$ has been also linked to the Edelstein effect \cite{Chen2024}, which causes a significant phase shift in the current-phase relationship of the junctions.

In a broad sense, the factors leading to inversion symmetry breaking are closely linked to the orbital properties, influencing the electronic structure near the Fermi level, encoded into the orbital Rashba interaction. Subsequently, the atomic spin-orbit coupling induces spin splitting via an effective spin Rashba coupling that arises as a combination of atomic spin-orbit coupling, orbital Rashba interaction and crystalline field potential. Although the orbital moment is expected to have a greater impact than spin polarization in the Edelstein effect, the exact influence of atomic spin-orbit coupling on the relationship between spin and orbital Edelstein effects in spin-orbit coupled superconductors is not yet fully settled.
Specifically, the atomic spin-orbit interaction $\lambda_{\text{}}$ 
is known to result in electronic states where the orbital angular momentum amplitude $\bm{{L}}$ is aligned or antialigned with the electron spin polarization $\bm{{s}}$ based on the sign of $\lambda$. However, although the spin-orbit coupling tends to favor configurations with parallel or antiparallel spin and orbital angular momentum, this characteristic may not necessarily carry over to the Edelstein spin and orbital polarization induced by the supercurrent. 

In this paper, we investigate the relationship between spin and orbital Edelstein effect in 2D multiorbital noncentrosymmetric superconductors by analyzing the impact and the role of the atomic spin-orbit coupling's sign and amplitude. For this purpose, we utilize a multiorbital model with $L=1$ orbital and $s=1/2$ electronic states at the Fermi level. Our results indicate that the relationship between the spin and orbital moments driven by the supercurrent is typically unrelated to the sign of the atomic spin-orbit coupling, and instead relies on the number of bands present at the Fermi level and the orbital properties of those bands.
To assess the character of the spin and orbital Edelstein effect we investigate its dependence in momentum space 
and focus on the role of avoiding crossings in setting out the size and sign of the induced magnetization. Remarkably, we find that the sign change of the orbital Edelstein effect occurring nearby the point of avoiding crossing bands is robust to a modification of the atomic spin-orbit amplitude. We also demonstrate that the spin Edelstein effect is typically one order of magnitude smaller than the orbital Edelstein one. However, for specific ranges of electron filling with pair of bands at the Fermi level that are marked by orbitals having equal mirror parity, we find that the spin Edelstein effect can be significantly enhanced by the increase of the atomic spin-orbit coupling. These outcomes indicate that the relation between the spin and orbital Edelstein effect can be tuned by a suitable control of the orbital degrees of freedom in noncentrosymmetric materials.

The paper is organized as follows. In Sec. II, we describe the model and the methodology. 
In Sec. III, we discuss the behavior of the spin and orbital moment in the normal state.  
In Sec. IV we show behavior of the spin and orbital Edelstein effect by considering the momentum resolved spin and orbital moments as a function of the atomic spin-orbit coupling and electron filling. The conclusions are reported in Sec. V. In the appendices we provide details about the spin-orbital character of the electronic states, the model Hamiltonian and the linear response of the Edelstein effect, as well as the derivation of effective down-folded models for the description of the Edelstein susceptibility in a 
spin-orbital coupled manifold with reduced dimension for various physical configurations.

\section{Model and Methodology}
In this section we introduce the model and the methodology which has been employed to investigate the spin and orbital Edelstein effect in 2D noncentrosymmetric superconductors with conventional $s$-wave spin-singlet pairing symmetry and equipped with multiple orbitals configurations at the Fermi level.
We analyze a two-dimensional electronic system with a local number of orbital degrees of freedom, which enables the inclusion of inversion symmetry breaking via the orbital Rashba interaction. In this system, three bands emerge from atomic orbitals within the $L=1$ angular momentum subspace, such as $p$ orbitals or $d_{\gamma}$ configurations where $\gamma$ can take on the values of $(yz,zx,xy)$. We specifically concentrate on $d$ orbitals within a square lattice geometry.
The normal state Hamiltonian is described by 
\begin{align}
H_N(\bm{k})=&\left[\begin{array}{ccc} \varepsilon_{yz} & 0 & -i \alpha \sin(k_x) \\  0 & \varepsilon_{zx} &  -i  \alpha \sin(k_y) \\  i \alpha \sin(k_x) &  i  \alpha \sin(k_y) & \varepsilon_{xy} \end{array} \right]\tilde{\sigma}_0\nonumber\\
&+\lambda_{\mathrm{}}\hat{\bm{L}} \cdot \tilde{\bm{s}}, 
\end{align}
\noindent 
where $\hat{L}_a$ are the orbital angular momentum operator described by
\begin{align}
\label{eq:angular}
&\hat{L}_{x}=\left[\begin{array}{ccc} 0 & 0 & 0  \\  0 & 0 & i  \\  0 &  -i & 0 \end{array} \right],
\hat{L}_{y}=\left[\begin{array}{ccc} 0 & 0 & -i \\  0 & 0 & 0  \\  i &  0 & 0 \end{array} \right],\nonumber\\
&\hat{L}_{z}=\left[\begin{array}{ccc} 0 & -i & 0 \\  i & 0 & 0  \\  0 &  0 & 0 \end{array} \right]
\end{align}
$\tilde{\sigma}_{a}$ are the Pauli matrices (with $a=x,y,z$),  and $\tilde{s}_a=\frac{1}{2}\tilde{\sigma}_a$ is the spin density operator, with
$\tilde{\sigma}_0$ being the unit matrix in the spin space and 
$\varepsilon_{yz}$, $\varepsilon_{zx}$, and $\varepsilon_{xy}$ are the diagonal components of the orbital dependent dispersion, which are given by
\begin{align}
\varepsilon_{yz}&=-2t^x_x\cos(k_x)-2t^y_x\cos(k_y)-\delta_x-\mu\\
\varepsilon_{zx}&=-2t^x_y\cos(k_x)-2t^y_y\cos(k_y)-\delta_y-\mu\\
\varepsilon_{xy}&=-2t^x_z\cos(k_x)-2t^y_z\cos(k_y)-\delta_z-\mu.
\end{align}
For clarity and convenience we assume that the hopping amplitudes are given by $t^y_{x}=t^x_{y}=t^y_{z}=t^x_{z}=t$ and 
$t^y_{y}=t^x_{x}=t'$, where we set $t=1.0$ and $t'=0.4$.
$\delta_{x}=\delta_{y}=0$, $\delta_{z}=0.5$ are the crystalline fields amplitude and $\mu$ is the chemical potential.    
$\alpha=0.1$ is the amplitude of the orbital Rashba interaction, which is induced by inversion symmetry breaking and $\lambda$ defines the strength of the atomic spin-orbit coupling. This choice of the electronic parameters is suitable to describe a distinct regime where the crystalline field potential is of the order of the hopping amplitude. This for instance applies to perovskite oxides materials.  

To describe the superconducting state, we decouple the s-wave spin-singlet pairing interaction so that one can write the Bogoliubov-de Gennes Hamiltonian within the particle-hole representation as:  
\begin{align}
  H=\frac{1}{2} \sum_{\bm{k}}\Psi^\dag_{\bm{k}} 
  H_{\bm{k}}\Psi_{\bm{k}},
\end{align}
with
\begin{align}
    H_{\bm{k}}=\left[\begin{array}{cc} 
    H_N(\bm{k}) & \mathcal{D}   \\
    \mathcal{D}^\dag & -H^*_N(-\bm{k})  
    \end{array} \right]
\end{align}
and
\begin{align}
    &\Psi^\dag_{\bm{k}}=[C^\dag_{\bm{k}},C^\mathrm{}_{\bm{k}}]\\
    &C^\dag_{\bm{k}}=[
  c^\dag_{\bm{k}\uparrow yz}
  c^\dag_{\bm{k}\uparrow zx}
  c^\dag_{\bm{k}\uparrow xy}
  c^\dag_{\bm{k}\downarrow yz}
  c^\dag_{\bm{k}\downarrow zx}
  c^\dag_{\bm{k}\downarrow xy}]
\end{align}
where $c_{\bm{k},s,\chi}$ is the creation operator of the electron with momentum $\bm{k}$, spin $s$, and orbital $\chi=(yz,zx,xy)$ degree of freedom.
Specifically, within this representation the $s$-wave spin-singlet pair potential can be expressed as
\begin{align}
  \mathcal{D}=\left[\begin{array}{ccc} \Delta & 0 & 0  \\  0 & \Delta & 0  \\  0 &  0 & \Delta \end{array} \right] i\tilde{\sigma}_y 
\end{align}
where $\Delta$ is the amplitude of the pair potential.
Here we set $\Delta=0.05$ in unit of the hopping amplitude. A change in the amplitude of $\Delta$ does not alter qualitatively the outcomes of the analysis in the regime where there are no gapless modes induced by the supercurrent flow.

The superconducting current can be introduced in a standard way by considering a spatial variation of the phase as $e^{i\bm{q} \cdot \bm{r}}$ in the order parameter. By gauge transformation, as shown in Appendix B, the phase change yields the following crystal wave-vector shift in the momentum space so that the Hamiltonian in the presence of a nonvanishing supercurrent amplitude can be expressed as:
\begin{align}
\label{eq:h8}
        H_{\bm{k},\bm{q}}=\left[\begin{array}{cc} 
    H_N(\bm{k}+\bm{q}/2) & \mathcal{D}   \\
    \mathcal{D}^\dag & -H^*_N(-\bm{k}+\bm{q}/2)  
    \end{array} \right]
\end{align}
In this paper, we choose $x$ as the direction of the supercurrent, which means that $\bm{q}=(q_x,0,0)$. Here we set $q_x=0.02$. 
Then we can expect finite expectation value of orbital and spin angular momentum in the $y$ direction and 
we can calculate 
the orbital and spin expectation value following as  
\begin{align}
  <L_y>=\frac{T}{N}\sum_{i \omega_n \bm{k}} \mathrm{Tr}[{\mathcal{G}}_{i\omega_n}(\bm{k},\bm{q}) L_y] \label{eq:orbital_Edelstein}
\end{align}
\begin{align}
  <s_y>=\frac{T}{N}\sum_{i \omega_n \bm{k}} \mathrm{Tr}[{\mathcal{G}}_{i\omega_n}(\bm{k},\bm{q}) s_y] \label{eq:spin_Edelstein}
\end{align}
with 
\begin{align}
    {\mathcal{G}}_{i\omega_n}(\bm{k},\bm{q})=
    [i\omega_n-H(\bm{k},\bm{q})]^{-1},
\end{align}
where $ {\mathcal{G}}_{i\omega_n}(\bm{k},\bm{q})$ is the Green function and $\omega_n=\pi T (2n+1)$ indicates the Matsubara frequency for the temperature $T$ and $n$ being an integer number.
Here orbital and spin angular momentum operator are given by
\begin{align}
    L_y=\left[\begin{array}{cc} 
    \hat{L}_y\tilde{\sigma}_0 & 0   \\
    0 &  \hat{L}_y\tilde{\sigma}_0 
    \end{array} \right],\\
    s_y=\frac{1}{2}\left[\begin{array}{cc} 
    \hat{L}_0\tilde{\sigma}_y & 0   \\
    0 &  \hat{L}_0\tilde{\sigma}_y 
    \end{array} \right],
\end{align}
or equivalently $L_y=\hat{L}_y\tilde{\sigma}_0 \bar{\tau}_0$ and $s_y=\frac{1}{2} \hat{L}_0\tilde{\sigma}_y \bar{\tau}_0$, by introducting the identity matrix operator $\bar{\tau}_0$ in the particle-hole subspace.  
$\hat{L}_0$ is the unit matrix in the orbital space.

\begin{figure*}[htbp] 
\includegraphics[width=0.85\linewidth]{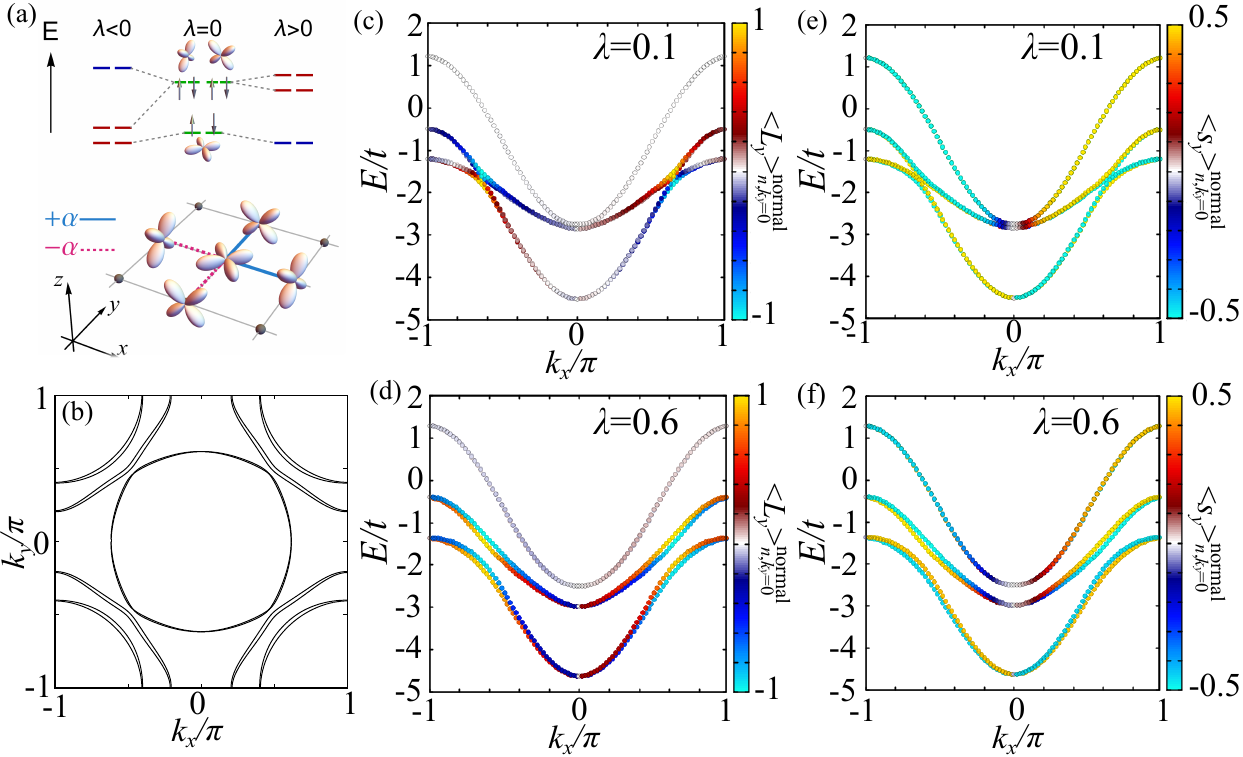}
\caption{(a) 
In the top part we show a sketch of the orbitals ${xy,xz,yz}$ spanning the $L=1$ manifold and of the spin-orbit coupled energy levels distribution at the $\Gamma$ point of the Brillouin zone. 
At zero spin-orbit coupling we schematically describe a configuration with tetragonal crystalline field potential ($\delta_z=0.5$) separating in energy the $xy$ orbital from the $xz,yz$ orbitals. Positive (negative) values of the spin-orbit coupling $\lambda$ lead to states with dominant $J=3/2$ (quartet in red) and $J=1/2$ (doublet in blue) character depending on the strength of the crystalline field splitting. Here, $J=L+s$ is the total angular momentum associated to electronic states with orbital moment $L=1$ and $s=1/2$. Due to the lack of rotational invariance for the crystalline field potential, quartets and doublets get mixed and they are not degenerate. 
In the bottom part of panel (a) we draw schematically the inter-orbital hybridization processes of electrons on nearest neighbor sites on a square lattice due to the orbital Rashba term $\alpha$. The orbital Rashba term breaks inversion symmetry, thus it is spatially odd (i.e. its amplitude reverses sign by inversion) and couples orbital states with different horizontal mirror parity to yield nonvanishing orbital angular momentum. (b) Fermi lines for a representative set of electronic parameters, i.e. chemical potential $\mu=0.0$, $\alpha=0.1$, $\delta_z=0.5$. Normal state electronic dispersion as a function of $k_x$ for $k_y=0$ and corresponding $k$-resolved expectation values of the orbital (c,d) and spin (e,f) moments at $\lambda=0.1$ and $\lambda=0.6$, respectively. 
}
\label{fig:figure1}
\end{figure*}



\section{Spin and orbital moment in the normal state}

Let us start by discussing the spin and orbital properties of the electronic states in the normal state of the examined system as a function of the spin-orbit coupling.
As mentioned in the previous section, due to the tetragonal crystalline potential, the $xy$ orbital configuration is energy split with respect to the $(xz,yz)$ states at the center of the Brillouin zone.
\textcolor{black}{In Fig. \ref{fig:figure1} (a) we show the schematic of the energy levels of the electronic states at the $\Gamma$ point in the Brillouin zone and the orbital dependent electron hybridization processes due to the orbital Rashba term that breaks inversion symmetry. For $\lambda=0$, $yz$ and $zx$ orbitals are degenerate and the $xy$ state is energy split. On the other hand, for nonvanishing $\lambda$, we have three Kramers pairs made of combination of $J=1/2$ doublet and $J=3/2$ quartets due to the lack of rotational symmetry. 
Here, $J=L+s$ is the total angular momentum associated to electronic states with orbital moment $L=1$ and $s=1/2$. The character of the electronic states at $\Gamma$ in terms of the spin and orbital moments can be tracked analytically (see Appendix A for details). For the choice made of the sign of $\lambda$ in the model we have that for positive (negative) $\lambda$, for zero crystalline field potential, the $J=1/2$ doublet is lower (higher) in energy as compared to the $J=3/2$ quartet, respectively.
}
The analysis is performed for a representative value of the crystal field potential. A change in the amplitude is not altering the qualitative overall trend of the examined spin and orbital Edelstein effect.  

Before considering the Edelstein effect in the superconducting state it is useful to discuss the spin and orbital moments distribution of the electronic states in momentum space.
In Fig. \ref{fig:figure1} we show a representative case of Fermi lines (b) and the electronic dispersion along a given direction in the Brillouin zone for a set of electronic parameters together with the corresponding evolution of the spin (e,f) and orbital (c,d) moments in the normal state. 
For clarity we select the $k_x$ direction in the Brillouin zone. Along this direction, the orbital Rashba coupling breaks the inversion symmetry by allowing for a nonvanishing $y$ component of the orbital moment $\langle \hat{L}_y \rangle$. We consider firstly two cases with different spin-orbit coupling for analyzing physical configurations with weak ($\lambda=0.1$) and moderate ($\lambda=0.6$) amplitude as compared to the crystalline and kinetic energy terms. The effect of the spin-orbit coupling can be directly accessed by inspection of the $\langle \hat{L}_y \rangle$ and $\langle \hat{s}_y \rangle$ profiles. 
We start observing that the orbital moment can have a large amplitude when electronic bands with different orbital character, with respect to the horizontal mirror parity, gets close in energy. This for instance occurs nearby the avoiding crossing since the hybridization of the orbitals and the orbital Rashba coupling are maximally effective into yielding a nonvanishing orbital moment (i.e. at $k_x \sim 0.6 \pi$ as in Fig. \ref{fig:figure1} (c), (d)). {\color{black}The occurrence of a sizable orbital moment is confined to the momenta that are nearby the avoiding crossing when the strength of the atomic spin-orbit coupling $\lambda$ is smaller than the orbital Rashba interaction.} Instead, the increase of $\lambda$ extends the effects to all the states across the Brillouin zone.

Since the orbital Rashba term acts as a $k$-dependent orbital field, the electronic states have an orbital moment with opposite sign through the avoiding crossing point. 
{\color{black} When considering the spin degrees of freedom, the orbital moment distribution nearby the avoiding crossing keeps exhibiting an opposite sign for the spin split bands. } However, the sign distribution depends on the strength of the atomic spin-orbit coupling. 
As expected, the combination of the atomic spin-orbit coupling and the orbital Rashba interaction act by removing the spin degeneracy at any $k$ point. Moreover, the spin-orbit coupling tends to unquench the orbital moment. Thus, a nonvanishing and sizable orbital moment can be also obtained away from the avoiding crossing point. Indeed, even the states close to $k_x \sim 0$ can acquire a sizable orbital component along the inversion broken direction even if the amplitude of the effective orbital Rashba field is negligible (Fig. \ref{fig:figure1} (c), (d)). We notice that the combination of the spin-orbit coupling and the orbital Rashba to get a nonvanishing orbital moment is not equally effective for all the bands. This behavior is a consequence of the crystalline anisotropy. Indeed, at $\lambda=0.1$ the lowest energy band has negligible orbital moment away from the avoiding crossing point. Instead, the highest energy bands have vanishing orbital moment along the $y$ direction at all momenta independently of the strength of the spin-orbit coupling. 


\begin{figure}[htbp] 
\includegraphics[width=1\linewidth]{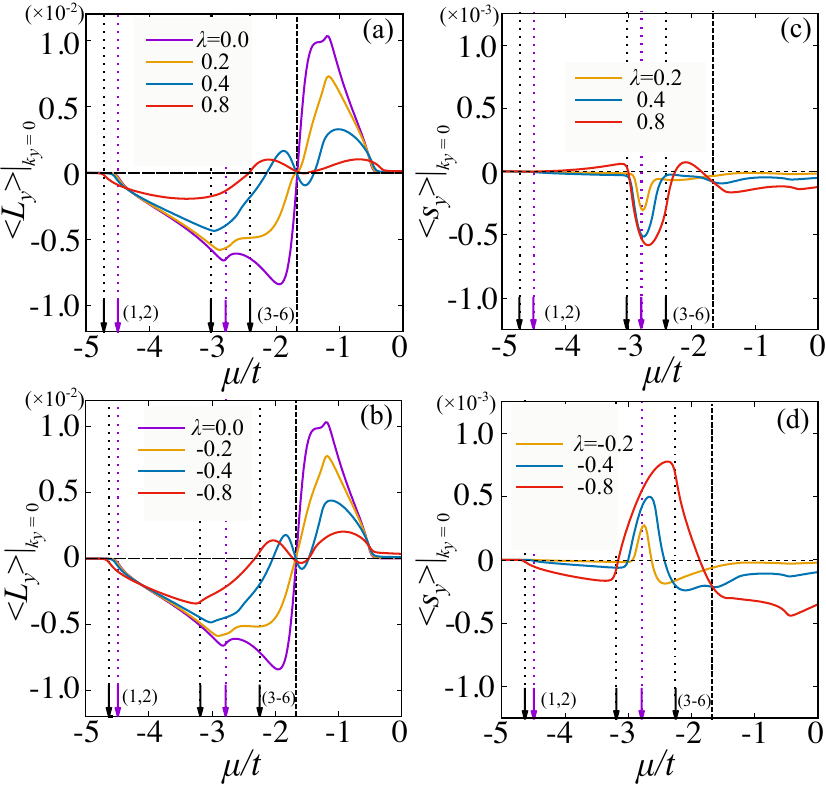}
\caption{
Energy dependent profile of the induced orbital Edelstein moment $L_y$ in the superconducting state, through the scan in the chemical potential $\mu$, for several values of the spin-orbit coupling $\lambda$ at $k_y/\pi=0.$: (a) $\lambda \geq 0$ and  (b) $\lambda \leq 0$.
Energy dependent profile of the induced spin Edelstein moment $s_y$ in the superconducting state, through the scan in $\mu$,  for several values of the spin-orbit coupling  at $k_y/\pi=0$ line: (c) $\lambda \geq 0$ and  (d) $\lambda \leq 0$.
The black dotted lines show the energy positions of the crossing points for $\lambda=0$ and $\alpha=0$. The purple (black) arrows indicate the energy positions of the bottom of the bands for $\lambda=0$ ($\lambda=0.8$ or $\lambda=-0.8$), respectively.  
}
\label{fig:figure3}
\end{figure}


\section{Results}

We now consider the spin and orbital moment that are induced by the application of a supercurrent for the examined multiorbital system by focusing on the role of the atomic spin-orbit coupling. The superconducting state is descrived by the spin-singlet $s$-wave pair potential with translational symmetry.
By symmetry, for the $C_{4v}$ point group,
the Edelstein effect can be expressed as
the wedge product of the polar vector related to the mirror symmetry-breaking direction ($\hat{z}$) and the supercurrent ($\mathbf{J}_c$), thus through the axial vector $\mathbf{M}\sim \hat{z} \times \mathbf{J}_c $ that is perpendicular to the applied supercurrent. For the examined system the magnetization is composed by the spin and orbital polarization. For convenience and clarity we consider the case of the supercurrent that is flowing along the $x$-direction. Hence, the induced Edelstein effect is composed by the spin and orbital moments along the $y$ direction only.

\subsection{Momentum resolved Edelstein effect}

\begin{figure}[htbp] 
\includegraphics[width=1\linewidth]{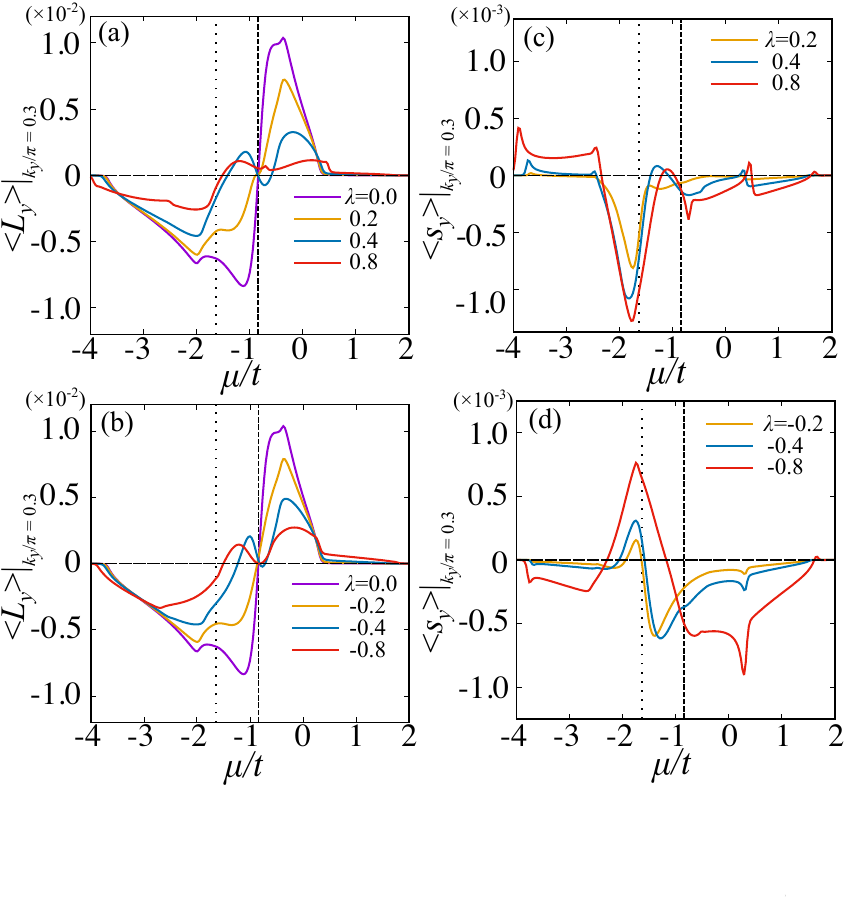}
\caption{Energy dependent profile of the induced orbital Edelstein moment $L_y$, through the scan in $\mu$, for several values of the spin-orbit coupling  at $k_y/\pi=0.3$ line: (a) $\lambda \geq 0$ and  (b) $\lambda \leq 0$ in the superconducting state.
Energy dependent profile, through the scan in $\mu$, of the induced spin Edelstein moment $s_y$ in the superconducting state for several values of the spin-orbit coupling  at $k_y/\pi=0.3$ line: (c) $\lambda \geq 0$ and  (d) $\lambda \leq 0$.
The thicker (thinner) black dotted line indicates the crossing point between $xy$ and $yz$ ($yz$ and $zx$) orbital for $\lambda=0$ and $\alpha=0$, respectively. 
}
\label{fig:figure4}
\end{figure}

We start by investigating the evolution of the Edelstein effect for a given cut within the Brillouin zone. This analysis helps to understand the overall behavior of the Edelstein effect. Indeed, we can focus on the amplitude of $\langle L_y \rangle$ and $\langle s_y \rangle$ by fixing the momentum transverse to the supercurrent flow direction. In Figs. \ref{fig:figure3} and \ref{fig:figure4} we report the energy-resolved spin and orbital polarization
in the superconducting state, for two representative cuts in the Brillouin zone at $k_y=0$ and $k_y=0.3 \pi$, respectively.

Let us start with the energy resolved profile of the orbital Edelstein effect at $k_y=0$. 
The energy profile of $\langle L_y \rangle$ evolves from negative to positive values with a sign change that occurs close to the avoiding crossing. Additionally, we observe that the orbital Edelstein moment, $\langle L_y \rangle$, is suppressed by the atomic spin-orbit coupling at all energies. We also notice that the overall sign of the orbital Edelstein effect is related to the orbital Rashba coupling and it can be reversed by changing the sign of $\alpha$. 
    On the other hand, the sign change of $\langle L_y \rangle$ as a function of the chemical potential   and the suppression of $\langle L_y \rangle$ can be understood by analysing the orbital Edelstein susceptibility $\chi_{yx}^{o}(k_x,k_y=0)$ within the $xy$ and $yz$ orbital space, to linear order in the momentum associated with the phase gradient of the supercurrent, by considering the energies close to the avoiding crossing point. 

{In Figs. \ref{fig:figure3} (a) and (b), the amplitude of $\langle L_y \rangle$ is vanishing at the point where the energies of the $xy$ and $yz$ bands cross, i.e. $\varepsilon_{xy}=\varepsilon_{yz}$.  
To understand the behavior of $\langle L_y \rangle$ close to this point we employ the analytical expression of the orbital Edelstein susceptibility $\chi_{yx}^{o}(k_x,k_y)$ obtained by neglecting the contributions of the $xz$ orbital component. The details of the derivation of $\chi_{yx}^{o}(k_x,k_y)$ for this case are reported in Appendix D.
The orbital Edelstein susceptibility can be expressed as:
\begin{align}
    \chi^o_{yx}(k_x,k_y)&=\frac{p}{p^2+(\alpha(k_x)-\lambda/2)^2}\nonumber \\&\times[(\alpha(k_x)-\lambda/2)\partial_{k_x}p -p\partial_{k_x}\alpha(k_x)]\nonumber\\
    &\times \frac{1}{{E_{1 }+{E_{2}}}} \left[  1-\frac{\varepsilon_{1 } \varepsilon_{2 } +\Delta^2}{E_{1 } E_{2 }}\right]\nonumber\\
   &+\frac{p}{p^2+(\alpha(k_x)+\lambda/2)^2}\nonumber \\&\times[(\alpha(k_x)+\lambda/2)\partial_{k_x}p -p\partial_{k_x}\alpha(k_x)]\nonumber\\
   &\times \frac{1}{{E_{3 }+{E_{4}}}} \left[  1-\frac{\varepsilon_{3 } \varepsilon_{4 } +\Delta^2}{E_{3 } E_{4 }}\right]
   \label{chio}
\end{align}
where the energy terms are given by 
\begin{align}
    & \varepsilon_1=q+R_-,\\
    & \varepsilon_2=q-R_-,\\
    & \varepsilon_3=q+R_+,\\
    & \varepsilon_4=q-R_+ \,,
\end{align}
with the coefficients $p$,\,$q$, $R_{\pm}$ and $ E_i$ that are given by the following expressions:
\begin{align}
    q=&(\varepsilon_{yz}+\varepsilon_{xy})/2,\\
    p=&(\varepsilon_{yz}-\varepsilon_{xy})/2,\\
    R_{\pm}=&\sqrt{p^2+(\alpha(k_x)\pm \lambda/2)^2},\\
    E_i=&\sqrt{\varepsilon_i^2+\Delta^2} ,\\
    \alpha(k_x)=&\alpha \sin(k_x).
\end{align}
Here, $\varepsilon_i$ with $i=1,..,4$ are obtained by diagonalizing the Hamiltonian of the normal state neglecting the $zx$ orbital contributions.
By inspection of Eq. (\ref{chio}), we can deduce the key features of the chemical potential dependence of the orbital Edelstein effect.
At the crossing point the orbital Edelstein susceptibility is given by: 
\begin{align}
    \chi^o_{yx}(k_x,k_y)&\sim\frac{p}{p^2+(\alpha(k_x)-\lambda/2)^2}(\alpha(k_x)-\lambda/2)\partial_{k_x}p \nonumber \\
    &\times \frac{1}{{E_{1 }+{E_{2}}}} \left[  1-\frac{\varepsilon_{1 } \varepsilon_{2 } +\Delta^2}{E_{1 } E_{2 }}\right]\nonumber\\
   &+\frac{p}{p^2+(\alpha(k_x)+\lambda/2)^2}(\alpha(k_x)+\lambda/2)\partial_{k_x}p \nonumber\\
   &\times\frac{1}{{E_{3 }+{E_{4}}}} \left[  1-\frac{\varepsilon_{3 } \varepsilon_{4 } +\Delta^2}{E_{3 } E_{4 }}\right].
   \label{chio_2}
\end{align}
Since  ${\chi^o_{yx}(k_x,k_y)} \sim p$  in Eq. (\ref{chio_2}), the sign change is robust close to the crossing point with $k_y=0$. Here, the slope of the $\langle L_y \rangle$ is related to the contribution $
{\chi^o_{yx}(k_x,k_y)}\sim\frac{1}{\alpha^2 -(\lambda/2)^2}$ as a function of $\lambda$ (see Appendix D).
For a broader energy region around the avoiding crossing point, the 
 Eq. (\ref{chio_2}) part is also relevant to grasp the behavior of the orbital Edelstein moment. Therefore, the induced orbital moment, $\langle L_y \rangle$ is negative (positive) for energies below (above) the crossing point, respectively.
 When $\lambda$ is the dominant energy scale we have that ${\chi^o_{yx}(k_x,k_y)}\sim\frac{1}{\alpha^2 -(\lambda/2)^2}$ and thus the orbital moment decreseas with the increase of the spin-orbit coupling.
The suppression of $\langle L_y \rangle$ can also be grasped by the distribution of the orbital moment in the normal state in Fig. \ref{fig:figure1} (c) (d). For smaller $\lambda$, we have that the spin split is small and the orbital moment has same sign for each pair of spin split bands in Fig. \ref{fig:figure1} (c). 
On the other hand, for larger $\lambda$, the spin split bands have orbital moment with opposite sign in Fig. \ref{fig:figure1} (d), which suggests that the contributions tend to cancel.
%
The analysis of the orbital Edelstein susceptibility also shows that a sign change in the spin-orbit coupling has almost no impact on the profile of the orbital Edelstein moment. This is confirmed by the numerical simulation presented in Fig. \ref{fig:figure3}. We also note that there are no distinct anomalies observed when the chemical potential crosses the values associated with the Lifshitz transition and the formation of electron pockets at the $\Gamma$ point, as illustrated by the dotted lines in Fig. \ref{fig:figure3}.
}

Let us now consider the behavior of the spin Edelstein effect along the $k_y=0$ direction.
The energy resolved evolution of the induced spin moment is reported in Fig. \ref{fig:figure3} (c),(d). We start by observing that the spin moment is vanishingly small for chemical potential such as only the first band is occupied. This regime corresponds to values of the chemical potential that are below $\sim -3 t$ both for positive and negative values of the spin-orbit coupling.
In this case, if one is down-folding the other electronic states which are well separated in energy, then we have a projected model with two spin configurations with an effective spin Rashba coupling, $\alpha_s^{xy}$, that is dependent on the orbital Rashba interaction, the spin-orbit coupling and the crystalline splitting. The details of the derivation are reported in Appendix E.  The amplitude of $\alpha_s^{xy}$ is proportional to $\frac{\alpha\lambda}  {\varepsilon_{yz(zx)}}$ \cite{Yanase2013}. 
Here, we observe that the effective model with only two spin-split bands cannot reproduce the behavior of the spin Edelstein effect for values of $k_y$ corresponding to time-reversal invariant momenta.
Indeed, as demonstrated in Appendix E, the spin Edelstein susceptibility for the downfolded effective spin-Rashba model obtained via the L\"owdin \cite{Lowdin1950} procedure is identically zero for $k_y=0,\pm \pi$. For this reason the leading order contribution of the induced spin moment corresponding to the lowest energy occupied band is not odd in $\lambda$ as one would have expected from the dependence of $\alpha_s^{xy}$. These observations
are validated by the direct calculation of the spin Edelstein moment (Fig. \ref{fig:figure3}(c),(d)), which remains unchanged in sign upon reversing the sign of $\lambda$. To understand the origin of the nonvanishing spin Edelstein effect in the regime of low electron filling with only two spin-split bands which are occupied, one has to explicitly take into account the multi-orbital degrees of freedom. 
In the Appendix F we have constructed an effective downfolded Hamiltonian obtained via the L\"owdin \cite{Lowdin1950} procedure which includes only two of the three orbital degrees of freedom. Then, the evaluation of the spin Edelstein susceptibility turns out to be nonvanishing also at the time-reversal points $k_y=0,\pm \pi$.
\\

We then move to the regime of having multiple bands occupied above the Lifshitz point close to $\mu  \sim -3 t$.
For this regime, when the bands with $xz$ and $yz$ character gets populated, the spin Edelstein effect becomes particularly enhanced by the spin-orbit coupling. This is a consequence of the fact that the $xz$ and $yz$ orbitals are close in energy and are not directly coupled through 
the orbital Rashba interaction because they have the same mirror parity \textcolor{black}{with respect to the horizontal plane}. Hence, you cannot induce a spin splitting among these orbital configurations by combining the orbital Rashba and the atomic spin-orbit interactions. For this reason, we have that the spin Edelstein is nonlinearly dependent on the spin-orbit coupling $\lambda$. For this type of bands we have derived an effective downfolded model obtained via the L\"owdin \cite{Lowdin1950} procedure and we have deduced an analytical expression for the spin Edelstein susceptibility (Appendix F). 
The obtained model is marked by two effective spin split bands which are orbitally entangled. When considering the spin Edelstein susceptibility, we have an analytical expression that while containing several terms, shows that the linear term in $\lambda$ is vanishing.
On the basis of this analysis, we can grasp some of the features of the spin Edelstein effect reported in Fig. \ref{fig:figure3}. 
We have an overall enhancement of the spin moment at all the values of the chemical potential that correspond to the occupation of the higher energy bands as a function of $\lambda$ due to the nonlinear dependence of the susceptibility. Furthermore, when we compare the scenarios with positive and negative spin-orbit coupling, we observe that the reversal does not occur across all the energy levels. The competition between the odd and even nonlinear contributions in $\lambda$ to the spin moment is clear from examining the expression for the spin Edelstein susceptibility (see Appendix F). The interplay of these contributions depends on the energy relationship between the relevant electronic states, which aligns with the observation that the energy-resolved $\langle s_y \rangle$ does not show a sign reversal for any value of the chemical potential (see Fig. \ref{fig:figure3} (c),(d) for $\mu \gtrapprox - 3 t$. Specifically, the odd parity cubic term in $\lambda$ is more effective at large $\lambda$ while for smaller values the quadratic term dominates. 
\\
Now, we can discuss the resulting relative sign of the induced spin and orbital Edelstein moments.
Regarding the orbital moment, we have demonstrated that its behavior is largely unaffected by the sign of $\lambda$. 
Nevertheless, we have also shown that spin-orbit coupling reduces the amplitude at all energy levels and can induce a sign change in $\langle L_y \rangle$ near the avoided crossing of bands that are directly coupled via orbital Rashba coupling (Fig. \ref{fig:figure3}(a),(b)).
Then, we have demonstrated that the spin Edelstein effect is enhanced when populating the bands with $xz$ and $yz$ character. In this regime the spin susceptibility has both even and odd nonlinear contributions in $\lambda$ that are however energy dependent. The resulting behavior is that a sign reversal of $\lambda$ leads to a trend with a reversal of the spin Edelstein moment only close to the Lifshitz point for the $xz,yz$ bands and for a sizable $\lambda$ as one can see in Fig. \ref{fig:figure3} (c),(d).  

The sign of the product of the spin and orbital Edelstein moment, ${\text{sign}}[\langle L_y \rangle \langle s_y \rangle]$, evaluated at $k_y=0$ shows that there is a reversal of the relative orientation only close to the Lifshitz point for the $xz,yz$ bands at $\mu \sim -3 t$. In the other range of energy the induced spin and orbital Edelstein moment are substantially unaffected by the reversal of sign of the spin-orbit coupling.

At this point, we can examine an alternative cut in the Brillouin zone to assess whether the behavior of the spin and orbital Edelstein moment is influenced by deviating from time-reversal invariant points, specifically at $k_y=0,\pm \pi$.
In Fig. \ref{fig:figure4}, we present the energy dependence of the spin and orbital Edelstein moments at $k_y/\pi=0.3$. 
There are a few key observations to note. Firstly, let us examine the induced orbital moment (Fig. \ref{fig:figure4} (a),(b)). The energy-dependent profile closely resembles that at $k_y = 0$, with the exception that the location of the sign change is modified. This is due to the fact that the crossing of the bands depends on $k_y$. As for the other cut at $k_y = 0$, we also observe that a change in the spin-orbit coupling results in a reduction of the orbital Edelstein moment. This phenomenon can be again captured by the behavior of the orbital Edelstein suscpetibility.

Then, let us consider the spin Edelstein response.
For the regime of chemical potential below $\mu \sim -2 t$ with only one pair of spin split bands we have that the spin Edelstein effect is sizable and it is substantially linear in $\lambda$. This is consistent with the spin Rashba model derived in Appendix E where the effective spin Rashba coupling scales linearly with $\lambda$. 
As for the previous case, the enhancement of the spin Edelstein effect is more effective when the $xz,yz$ bands get occupied.
On the basis of these results one can immediately deduce the behavior for the relative sign of the spin and orbital Edelstein effect. Overall, the reversal of $\lambda$ does not induce a reversal of the sign of the $L s$ product except for low electron filling when the lowest energy pair of spin split band is occupied. 
The sign change of the spin Edelstein effect occurs for energies that correspond to the occupation of the intermediate pairs of bands with $xz,yz$ orbital character. For these configurations however, since the effective spin Rashba coupling is nonlinear in $\lambda$ there can be a sign reversal when the strength of $\lambda$ is sufficiently large.

\subsection{Spin and orbital Edelstein effect}
\begin{figure}[htbp] 
\includegraphics[width=1\linewidth]{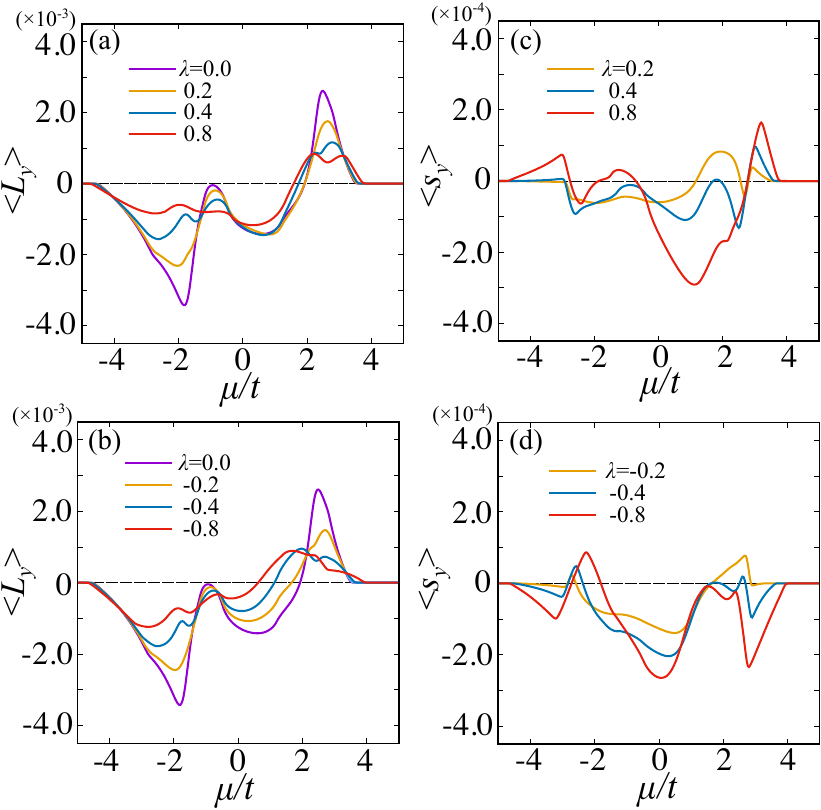}
\caption{Energy resolved spin and orbital Edelstein moment versus $\lambda$ for the examined noncentrosymmetric superconductor. $L_y$ expectation values integrated over the whole Brillouin zone and evaluated for different strength of the spin-orbit coupling: (a) $\lambda \geq 0$ and  (b) $\lambda \leq 0$.
Spin Edelstein effect as given by the $s_y$ amplitude of the induced spin moment for (c) $\lambda \geq 0$ and  (d) $\lambda \leq 0$.
}
\label{fig:figure5}
\end{figure}

After having evaluated the momentum resolved Edelstein effect, we consider the total orbital and spin responses obtained by averaging across the entire Brillouin zone in Fig. \ref{fig:figure5} (a)-(b) and (c)-(d), respectively.
We start by discussing the orbital Edelstein effect. 
One can observe that in the energy profile there are peak structures near $\mu/t\sim \pm3$ and a broad negative 
peak near $\mu/t\sim$ 1. Moreover, when comparing the total orbital Edelestein moment with that of the momemtum resolved configuration, we see that 
There is an amplitude suppression of about $\sim 1/10$ around the broad peak. 
As for the momentum resolved case, the increase of the spin-orbit coupling leads to a reduction of the orbital Edelstein amplitude.
\\
The momemtum dependence orbital Edelstein effect can help understanding the resulting hallmarks of the average orbital Edelstein moment. 
The structures depicted in Fig. \ref{fig:figure5} can be indeed directly interpreted through the energy profile associated with the momentum-resolved Edelstein effect. For example, when examining the momentum-resolved orbital Edelstein effect, we find that for any specific transverse momentum ($k_y$), the profile varies with the chemical potential, consistently displaying a sign change at a critical value of $\mu$. This value depends on the given $k_y$ and refers to the energy position of the avoided crossing point in the electronic structure. Moreover, the amplitude of the orbital Edelstein effect is amplified near the avoided crossing, since there the orbital Rashba yields a maximal value of the orbital moment, and, as expected, the orbital moment diminishes moving to energies corresponding to fully empty or fully occupied bands.
To comprehend the overall profile of the orbital Edelstein effect, it is important to note that the point at which $L_y$ changes sign is influenced by $k_y$, shifting from approximately $\mu \sim -2 t$ to $\mu \sim 2.5 t$ as $k_y$ is varied from the center of the Brilluion zone ($k_y$=0) to zone boundary ($k_y$$=\pi$).
Hence, one can observe that the position where $\langle L_y \rangle$ changes sign depends on $k_y$ in a way that it shifts from $\mu \sim -2 t$ to $\mu \sim 2 t$ when $k_y$ is varied from 0 to $\pi$. Taking into account this aspect and the fact that both the profile and the energy widths where the orbital Edelstein moment exhibits a negative or positive value are not significantly affected by $k_y$, one can observe that the average over $k_y$ will lead to a total orbital Edelstein moment which is peaked at $\mu \sim \pm 2.5 t$. The peaks are the result of the convolution of the various $k_y$ contributions close to the avoiding crossings.  
To make more evident the trend of $\langle L_y \rangle$ and the shift of the point where the orbital Edelstein effect is vanishing one can follow the evolution for several values of $k_y/\pi$ (Fig. 8 of the Appendix). The overall trend is then evident by comparing the momentum resolved orbital Edelstein effect from small momentum to large momentum at zone boundary.
For instance, for $k_y/\pi=0.7$ one can see that $\langle L_y \rangle$ starts to be nonvanishing for values of $\mu > -2 t$.  This implies that for those or larger momenta there will be a dominant negative contribution in the energy window among $-2 t$ and $1.5 t$. This trend also accounts for the broad negative distribution near $\pm t$. 
Hence, one can conclude that for moments up to roughly $k_y/\pi=0.5$ the average will provide a cancellation of the positive contributions above the critical $\mu$ associated with the avoiding crossing and the peak position at about $-2 t$. While for momenta about larger than $k_y/\pi=0.5$ the average builds up the broad negative part and the positive peak at about $1.5 t$.
The sign of the orbital Edelstein effect remains mostly unchanged despite the reversal of the spin-orbit coupling sign and the primary effect of atomic spin-orbit coupling on the orbital Edelstein effect is to diminish the strength of the induced orbital moment as shown in Fig. \ref{fig:figure5}(a) and (b). These features are consistent with the outcomes of the obital Edelstein moment on small $k_y$ lines.   
\\
Regarding the suppression of the amplitude of the total Edelstein orbital moment compared to the momentum resolved one, it arises because not all $k$ orbital Edelstein components contribute at a given energy, and there are cancellation effects when averaging the positive and negative contributions across different momentum values.
Indeed, for energies near $-2t$, only small $k_y$ momenta play a significant role, whereas in the intermediate energy range between $-t$ and $t$, momenta near 0.5$\pi$ dominate the negative response, and the positive orbital Edelstein effect at high energies arises from momenta close to the zone boundary.
Therefore, one could consider designing a quasi 2D superconducting strip with a specific lateral size that allows only certain bands to be occupied for a particular electron filling. This would result in only a few channels contributing to the orbital Edelstein moment, thereby minimizing cancellation effects. 
One alternative solution to bypass the suppression is to achieve an electronic dispersion in the transverse direction that is significantly smaller than that along the direction of the supercurrent flow. This approach would ensure that the location of the avoided crossing is less dependent on $k_y$, thereby diminishing the cancellation effects caused by the momentum-dependent energy shifts of the orbital Edelstein moment.

When examining the spin Edelstein effect, we note that an increase in $\lambda$ generally leads to an enhancement of the spin moment independently on its sign, with the exception of a few small energy regions (Fig. \ref{fig:figure5}(c),(d)). 
This behavior generally corresponds with the property  of the spin Edelstein effect at given $k_y$ close to the center of the Brillouin zone, which, however, can display either a linear or nonlinear relationship with $\lambda$, depending on the orbital properties of the electronic states at the Fermi level. 
Our analysis shows that the enhancement of the spin Edelstein effect is significantly greater for superconducting states characterized by orbital configurations that are not directly influenced by orbital Rashba coupling. For the model system under examination, this situation corresponds to electronic states with a predominant $xz$ and $yz$ orbital character when the Fermi level is approximately above $-2t$ as wee can see in Fig. \ref{fig:figure5} (c) and (d).

Concerning the behavior of the spin Edelstein effect when the sign of $\lambda$ is reversed, as well as the associated behavior of the $L s$ product, we observe that a correlation exists only in certain specific regions of the energy parameter space. Specifically, it is the case of a singly occupied spin split band (i.e. for $\mu$ below about $-2 t$ and above about $3 t$) that has a tendency to exhibit a reversing of the sign of the $L s$ product.
This behavior can be accounted by the effective spin Rashba coupling as due to the combination of the spin-orbit coupling $\lambda$ and the orbital Rashba coupling. 
This trend extends over the energy range when the amplitude of the spin-orbit coupling grows. 

The spin Edelstein effect is one order of magnitude smaller than the orbital Edelstein one 
even if the absolute value of spin-orbit coupling $|\lambda|$ becomes comparable to the hopping amplitude $t$. This tendency is consistent with the momentum resolved spin Edelstein effect for any $k_y$. 
By examining the momentum-resolved spin Edelstein effect, we find that at energy levels where all three orbitals are hybridized and occupied, there is a significant spin contribution. This holds not only for small $k_y$ but also for $k_y$ close to the zone boundary (see Fig. \ref{fig:figure6} (c) (d) in Appendix G). 

The amplitude of the spin and orbital Edelstein moments has been evaluated for a supercurrent that is roughly one-tenth of the critical amplitude. This estimate has been based on the employed value of the momentum $q$ (i.e. $q=0.02$) in the phase gradient. The order parameter is suppressed at about $q \sim q_c$ with $q_c \sim 0.2$, thus $q_c$ corresponds to a maximal critical supercurrent that the superconductor can sustain. 
Consequently, we have assumed an orbital Rashba coupling value of about 20 meV, given that the typical value of $t$ falls within the hundreds of meV range for transition metal or oxides superconductors.
The orbital moment is measured in units of the Bohr magneton. Therefore, even without finely tuning the chemical potential, the induced orbital magnetization can reach approximately $10^{-3}$ Bohr magneton, which corresponds to magnetic dipolar fields on the order of hundreds of Gauss near the surface of the superconductor where supercurrents flow. Our research has shown that the orbital Edelstein moment is generally around an order of magnitude larger than the spin Edelstein moment. 



\section{Conclusions}

We have investigated the spin and orbital Edelstein effect in spin-singlet noncentrosymmetric superconductors characterized by electronic states that feature coupled spin and orbital moments. 
We have concentrated on the influence of atomic spin-orbit coupling in determining both the sign and magnitude of the induced spin and orbital moments when a supercurrent is applied. 
We demonstrate that, while the spin-orbit coupling leads to the expected electronic configurations in the normal state at the Fermi level, showing aligned (antialigned) angular momenta ($L$) and ($s$) corresponding to positive (negative) amplitudes of the ($L s$) product, the spin and orbital Edelstein effects do not behave similarly.
In fact, we show that the sign of the orbital Edelstein effect remains substantially unchanged despite variations in the sign and magnitude of the spin-orbit coupling. 
The amplitude of the induced orbital moment is significantly influenced by variations in the strength of the spin-orbit coupling, regardless of its sign.
Our analysis demonstrates that an increase in spin-orbit coupling typically leads to a decrease in the orbital Edelstein moment independently of the character of the electronic states. A hallmark of the orbital Edelstein effect is indicated by the sign change occurring near the avoided crossing point of electronic states, which can interact via the orbital Rashba effect and atomic spin-orbit coupling. We discover that the sign reversal of the orbital Edelstein effect near the avoided crossing remains largely unchanged despite variations in the sign and magnitude of the spin-orbit coupling. Additionally, we have demonstrated that, within an effective orbital projected description, the orbital Edelstein susceptibility is proportional to the energy band separation assessed at zero spin-orbit and orbital Rashba interactions. This result indicates that the sign reversal of the orbital Edelstein moment is fixed at a specific value of the chemical potential, which corresponds to the energy bands crossing. 
Consequently, this behavior enables a precise prediction of its position, regardless of the strength of the spin-orbit coupling and inversion-symmetry-breaking interactions.

The spin Edelstein effect demonstrates a distinctly different behavior when compared to the orbital Edelstein effect. We observe that the amplitude of the spin Edelstein moment generally rises with increasing strength of $\lambda$. However, this variation is influenced by the characteristics of the bands and the character of the involved orbitals.
In fact, the spin Edelstein effect is relatively weak for a single spin-split band and only becomes significant under conditions of strong spin-orbit coupling. Instead, for multiple bands occurring at the Fermi level the overall spin moment is large even in the regime of weak spin-orbit coupling and can get significantly enhanced when moving to a regime of strong spin-orbit interaction. 
Let us now discuss how the sign of the spin generated by the Edelstein effect can vary. This sign determines the direction of the spin polarization (i.e., whether more spins of one type are aligned compared to the other).
Our results show that the sign of the Edelstein effect related to spin is generally influenced by the orbital characteristics of the bands. In particular, the sign can be inverted in a more controllable manner by changing the sign of $\lambda$ when one pair of spin-split bands is occupied, as it occurs at low electron density for the examined multiorbital system.

We would like to highlight the relationship between the relative sizes of the spin and orbital Edelstein effects as the strength of the spin-orbit coupling is varied.
Our results suggest that while the orbital Edelstein effect is diminished due to spin-orbit interaction, the amplitude of the induced orbital moment remains an order of magnitude larger than that of the spin moment, even in the regime of strong spin-orbit coupling.

Finally, we point out that our results indicate that typically the spin-orbit coupling cannot decide the orientation of the spin moment induced by the supercurrent. This is quite different from what is typically expected in the relationship between the orbital Hall effect and the spin Hall effect \cite{Go2018}, where the spin Hall effect can occur in the same or opposite direction as the orbital Hall effect, depending on whether the $L s$ correlation of the electronic states is positive or negative.
Our results provide a clearcut guidance for designing spin-orbit driven magneto-electric Edelstein effects in noncentrosymmetric multiorbitals superconductors.
The uncovered spin and orbital Edelstein effects do not follow the spin-orbit driven $L s$ correlations. Therefore, one can effectively utilize superconducting materials made from different elements that exhibit varying strengths of spin-orbit coupling or are characterized by combinations of electronic states at the Fermi level with either low or high $J$, as these would not experience interference between the induced spin and orbital Edelstein effects. 

Concerning the design of materials with strong or weak spin-orbit coupling in principle it is quite straightforward because it relates to the atomic mass of the constituent elements. This implies that candidates superconductors with strong spin-orbit interaction link to elements as Ta, Bi, Nb, Pb, Rh, W, etc.
However, the combined role of electron correlations and crystalline field potential can also amplify the spin-orbit coupling thus opening other paths for materials designing with lighter elements \cite{Li2022}. 

To date experimental evidences of the Edelstein effects have been indirectly obtained by considering the impact of the induced magnetization on the superconducting transport properties and on the Josephson effect \cite{Senapati2023,Chen2024}.
However, on a general ground, the spin and orbital Edelstein effects can be directly detected using magnetic probes capable of measuring small magnetic moments, such as scanning SQUID \cite{Kirtley_2010}, nitrogen-vacancies \cite{Maletinsky2012} and magnetic exchange force microscopies \cite{Kaiser2007}, magneto-optic Kerr effect \cite{Qiu2000} and muon spectroscopies, among others.
One challenge here is that the magnetization is typically small in amplitude and can be spatially inhomogeneous. Moreover, one has to conduct in operando experiments with measurements that require magnetic detection while a supercurrent is flowing through the superconductor. 
In this framewortk, since the Edelstein effect depends on the lack of inversion symmetry, employing magnetic probes that are responsive to surface and interface phenomena can yield access to a broader variety of physical configurations. 

It is important to point that the orbital Edelstein moment typically exceeds the spin Edelstein moment, making it more detectable. This signifies that multiorbital noncentrosymmetric materials are particularly relevant for experimental investigations of this effect. Furthermore, the strength of the orbital Edelstein moment is directly related to the intensity of the orbital Rashba coupling, which correlates with the extent of inversion symmetry breaking. By applying an electrostatic potential or strain field, one can enhance these inversion symmetry breaking effects. Therefore, one can argue that utilizing these external influences could further strengthen the Edelstein effect.

In this context, our study indicates that when we interface superconducting materials with different atomic numbers or electronic states with unequal spin-orbit coupling strength, the spin Edelstein effect may tend to cancel out, while the orbital Edelstein effect remains robust, showing only a variation in amplitude across the interface.
This suggests that the induced magnetic moment at the interface of superconductors, marked by electronic states with strong and weak spin-orbit coupling, can be employed to single out the contribution of the orbital Edelstein effect with respect to the spin one.


\section*{Acknowledgments}
S.A. was supported by JSPS Grant Number 24KJ1221.
M.C. and M.T.M. acknowledge support from the EU’s Horizon 2020 research and innovation program under Grant Agreement No. 964398 (SUPERGATE). M.T.M.  acknowledges partial support by the Italian Ministry of Foreign Affairs and International Cooperation, grants KR23GR06 (MAP) e PGR12351 (ULTRAQMAT). This research has been supported by PNRR MUR project PE0000023-NQSTI.
Y.\ T.\ acknowledges financial support from JSPS with Grants-in-Aid
for Scientific Research (KAKENHI Grants Nos.\ 23K17668, 24K00583, and 24K00556).

\begin{figure*}[htbp] 
\includegraphics[width=1\linewidth]{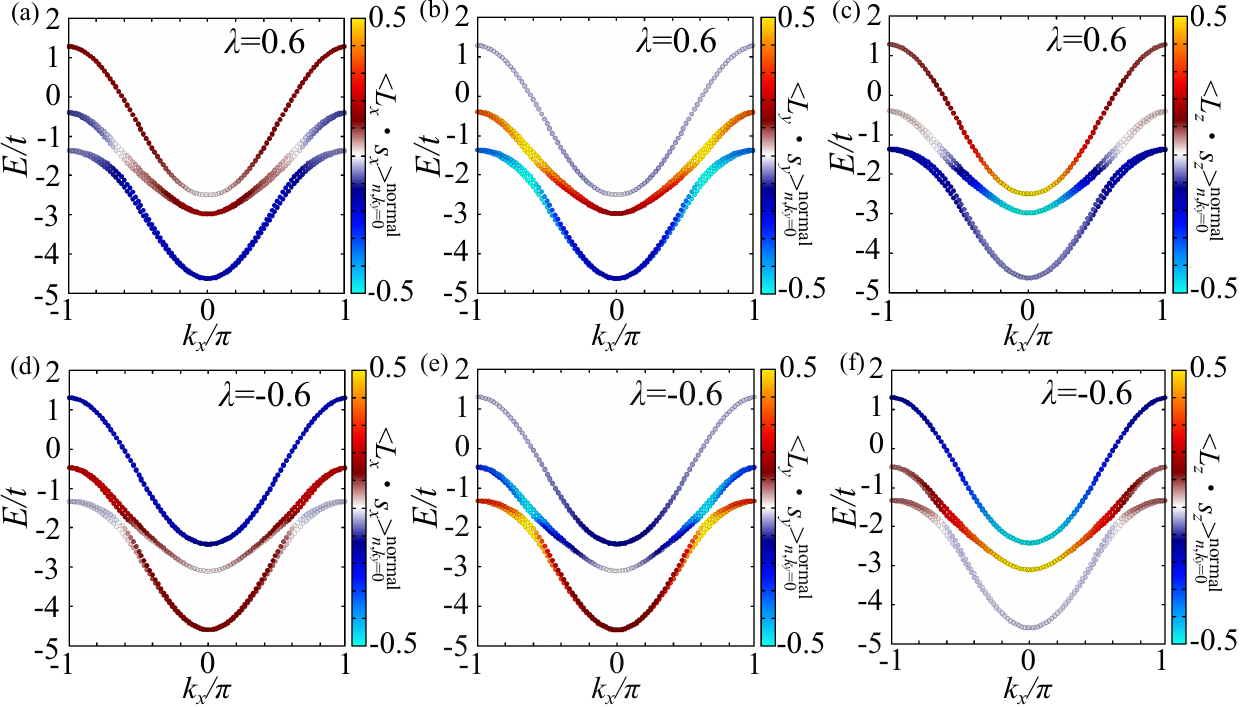}
\caption{Energy resolved expectation value of $L_a\cdot s_a (a=x,y,z)$ in the normal state. (a) $a=x$, (b) $a=y$, (c) $a=z$ for $\lambda=0.6$.
(d) $a=x$, (e) $a=y$, (f) $a=z$ for $\lambda=-0.6$. In these figures, we use $\alpha=0.1 t$, $\Delta=0.05 t$, and zero temperature,$T=0$.}
\label{fig:figure2}
\end{figure*}

\section*{Appendix A: Relation between spin and orbital moments in the normal state}

\textcolor{black}{Here, we discuss the role of the spin-orbit coupling in setting out the spin and orbital moments distribution for electronic states nearby the time reversal invariant points where the orbital Rashba interaction is vanishing (e.g. at $\Gamma=(0,0)$, as well as at $X=(\pm \pi,0)$ or $Y=(0,\pm \pi)$ and $M=(\pm \pi,\pm \pi)$ points). For these values of the momentum the electronic state are Kramers degenerate pairs which are split by the spin-orbit coupling ($\lambda$) and the tetragonal crystalline field potential ($\delta_t$). The latter includes the kinetic contribution too. The Kramers degenerate eigenvalues ($E_i$ with $i=1,2,3$) of the Hamiltonian $H_N(\bm{k}=0)$ have the following analytical expression:
\begin{eqnarray*}
    E_1&&=\frac{\lambda}{2},\nonumber\\
    E_2&&=\frac{1}{2}(\delta_t-\frac{\lambda}{2}-\sqrt{{\delta_t}^2-\lambda {\delta_t}  +\frac{9}{4} \lambda^2}),\nonumber\\
    E_3&&=\frac{1}{2}(\delta_t-\frac{\lambda}{2}+\sqrt{{\delta_t}^2-\lambda {\delta_t} +\frac{9}{4} \lambda^2}) \,
\end{eqnarray*}
\noindent with $\delta_t=\varepsilon_{xy}-\varepsilon_{yz(zx)}$ being the energy difference between $xy$ and $yz(zx)$ orbital at the $\Gamma$ point assuming that the zero is set at the energy of the $yz(zx)$ states. }

\textcolor{black}{The states $\psi_{1,\pm}$ associated to the energy $E_1$ correspond to the configurations with the highest $J_z=\pm 3/2$ projections of the $J=3/2$ quartet. Hence, depending on the sign of $\lambda$ being positive (negative) the configurations $\psi_{1,\pm}$ are the {highest(lowest)} in energy within the multiplet manifold at the $\Gamma$ point. 
For this configurations the spin and orbital moments have collinear direction with the same orientation. 
Since the configuration $\psi_{1,\pm}$ has the maximum allowable orbital moment projection, it is clear that the expectation value of the $L_y$ orbital component is small as demonstrated in Figs. \ref{fig:figure1} (c,d).
Instead, the energy splitting $\Delta E_{23}=E_2-E_3$ is always negative independently on the values assumed by $\delta_t$ and $\lambda$ away from the limits $\delta_t=\lambda=0$. This implies that the Kramers states with energy $E_2$ are placed in the middle of the manifold. From the inspection of the eigenvector of $H$, the state with energy $E_2$ has a character that is a mixing of the $J=3/2$ quartet and $J=1/2$ doublet with a dominant component from the low projections configurations of the $J=3/2$ multiplet. Finally the state $E_3$ has mostly a $J=1/2$ character, thus the spin and orbital momenta are anti-aligned.
Therefore, when $\lambda$ is positive, the lowest energy pair of bands exhibit antialigned spin and orbital moments because of the dominant $J=1/2$ character, as illustrated in Figs. \ref{fig:figure1} (c,d,e,f). Conversely, the spin and orbital moments of the second and third group of bands at higher energies have the same orientation, attributed to the $J=3/2$ character of the spin-orbit coupled states.} 

\textcolor{black}{The trend about the relation between the spin and orbital moment in the normal state can be further tracked by investigating the expectation values of the products of the spin-orbit coupling along the $x,y,z$ symmetry directions as given by $\langle L_x s_x \rangle$, $\langle L_y s_y \rangle$ and $\langle L_z s_z \rangle$. Figure \ref{fig:figure2} shows the different projections of the spin-orbit coupling in a representative case, taking into account the opposite sign of the spin-orbit interaction.
As discussed above, for the expression of the Hamiltonian, a positive (negative) $\lambda$ value yields a hierarchy of the multiplet with the $J=1/2$ doublet that is lower (higher) in energy than the $J=3/2$ quartet. Consequently, the spin and orbital moments along a given direction are preferentially anti-aligned (aligned) for the $J=1/2$ doublet ($J=3/2$ quartet), respectively. This behavior is investigated in Fig. \ref{fig:figure2} along a given direction in momentum space.
Let us start discussing the case of positive $\lambda$. The overall trend indicates that the lowest (highest) energy bands have spin and orbital moments that are anti-aligned (aligned), respectively, i.e. the expectation value of  $\langle L_i s_i \rangle$ is negative (positive) for $i=x,y,z$. Instead, the intermediate energy has a profile that is $k$- and orientation dependent. For the orientations that are perpendicular to the $y$ direction, the spin and orbital moment are parallel aligned before the position of the avoiding crossing and then there is a reversal of the orientation. In the $y$-direction, the polarization effect of the orbital Rashba coupling dictates that the relative orientation aligns with the sign of the spin-orbit coupling across all momentum values.
Hence, since we are interested in assessing the spin and orbital Edelstein effect, we focus on the behavior of the $y$-component of the spin and orbital moment. As illustrated in Fig. \ref{fig:figure2}, reversing the sign of the spin-orbit interaction leads to a reversal in the relative orientation of the spin and orbital moments across all momenta and energy levels.}

\section*{Appendix B: Hamiltonian and supercurrent}

In this appendix we show the gauge transformation on particle and hole states, which map the position dependence of the order parameter,  
due to the presence of the current bias, in a momentum shift of opposite sign for particle and holes, leading to the Hamiltonian \eqref{eq:h8}.
We start by writing the BdG Hamiltonian in real space
\begin{align}
H^{\mathrm{ph}}&=\frac{1}{2}\int d\bm{r} \int d\bm{r}'  
    \Psi^{\dag}_{\bm{r}} H_{\bm{r}-\bm{r}'} \Psi_{\bm{r}'}\\
    H_{\bm{r}-\bm{r}'}&=
           \left[  
      \begin{array}{cc}
      h_{\bm{r}-\bm{r}'} & \Delta\delta_{\bm{r},\bm{r}'}\mathrm{e}^{2i\bm{q}\cdot\bm{r}} \\
      \Delta\delta_{\bm{r},\bm{r}'}\mathrm{e}^{-2i\bm{q}\cdot\bm{r}}& -h^*_{\bm{r}-\bm{r}'}
      \end{array}
           \right]\\
        \Psi^\dag_{\bm{r}}&=[c^\dag_{\bm{r}},c_{\bm{r}}],
\end{align}
where $c_{\bm{r}}$ is the annihilation operator and $h_{\bm{r}-\bm{r}'}$ is the normal part of the Hamiltonian. As mentioned above, the order parameter position position-dependent phase, 
$\phi(\bm{r})=2 i \bm{q}\cdot\bm{r}$, is due to the current bias.
Here we do not include the spin degrees of freedom, since they are not affected by the gauge transformation.
We now perform the following unitary transformation
\begin{align}
    H'_{\bm{r}-\bm{r}'}&=U_{\mathrm{sc}}(\bm{r})H_{\bm{r}-\bm{r}'}U^\dag_{\mathrm{sc}}(\bm{r})\nonumber\\       
U_{\mathrm{sc}}(\bm{r})&=\left[  
      \begin{array}{cc}
      \mathrm{e}^{-i\bm{q}\cdot\bm{r}}& 0 \\
      0& \mathrm{e}^{i\bm{q}\cdot\bm{r}} 
      \end{array}
           \right],
\end{align}
and for the basis we have
\begin{align}
    \Psi^\dag_{\bm{r}}\xrightarrow{}\Psi^{'\dag}_{\bm{r}}=[c'^{\dag}_{\bm{r}},c'_{\bm{r}}]=[\mathrm{e}^{i\bm{q}\cdot\bm{r}}c^\dag_{\bm{r}},\mathrm{e}^{-i\bm{q}\cdot\bm{r}}c_{\bm{r}}].
\end{align}
Next, using the Fourier transform, we rewrite the Hamiltonian in momentum space
Explicitly, for the electron part,
\begin{align}
    &\frac{1}{V^2}\int \int d\bm{r} d\bm{r}' \delta_{\bm{r}-\bm{r}',\bm{n}}c'^{\dag}_{\bm{r}}c'_{\bm{r}'}\nonumber\\
    &=\frac{1}{V^2}
     \int \int d\bm{r} d\bm{r}' \delta_{\bm{r}-\bm{r}',\bm{n}}\sum_{\bm{k}}\mathrm{e}^{i(\bm{k}+\bm{q})\cdot\bm{r}}c_{\bm{k}}^{\dag}
     \sum_{\bm{k}'}\mathrm{e}^{-i(\bm{k}'+\bm{q})\cdot\bm{r}'}c_{\bm{k}'}
     \nonumber\\
     &=\frac{1}{V}
     \int d\bm{r}  \delta_{\bm{r}-\bm{r}',\bm{n}}\sum_{\bm{k}}\mathrm{e}^{i(\bm{k}+\bm{q})\cdot\bm{r}}c_{\bm{k}}^{\dag}
     \sum_{\bm{k}'}\mathrm{e}^{-i(\bm{k}'+\bm{q})\cdot(\bm{r}-\bm{n})}c_{\bm{k}'}
     \nonumber\\
     &=\frac{1}{V}\int d\bm{r} \sum_{\bm{k}}
     \sum_{\bm{k}'}\mathrm{e}^{i(\bm{k}-\bm{k}')\cdot\bm{r}}\mathrm{e}^{i(\bm{k}'+\bm{q})\cdot\bm{n}} c_{\bm{k}}^{\dag} c_{\bm{k}'}
     \nonumber\\
          &=\sum_{\bm{k}}
\mathrm{e}^{i(\bm{k}+\bm{q})\cdot\bm{n}} c_{\bm{k}}^{\dag} c_{\bm{k}},
\end{align}
while for the hole part,
\begin{align}
    &\frac{1}{V^2}\int \int d\bm{r} d\bm{r}' \delta_{\bm{r}-\bm{r}',\bm{n}}c'_{\bm{r}}c'^{\dag}_{\bm{r}'}\nonumber\\
    &=\frac{1}{V^2}
     \int \int d\bm{r} d\bm{r}' \delta_{\bm{r}-\bm{r}',\bm{n}}\sum_{\bm{k}}\mathrm{e}^{i(\bm{k}-\bm{q})\cdot\bm{r}}c_{-\bm{k}}
     \sum_{\bm{k}'}\mathrm{e}^{-i(\bm{k}'-\bm{q})\cdot\bm{r}'}c^{\dag}_{-\bm{k}'}
     \nonumber\\
     &=\frac{1}{V}
     \int d\bm{r}  \delta_{\bm{r}-\bm{r}',\bm{n}}\sum_{\bm{k}}\mathrm{e}^{i(\bm{k}-\bm{q})\cdot\bm{r}}c_{-\bm{k}}
     \sum_{\bm{k}'}\mathrm{e}^{-i(\bm{k}'-\bm{q})\cdot(\bm{r}-\bm{n})}c^{\dag}_{-\bm{k}'}
     \nonumber\\
     &=\frac{1}{V}\int d\bm{r} \sum_{\bm{k}}
     \sum_{\bm{k}'}\mathrm{e}^{i(\bm{k}-\bm{k}')\cdot\bm{r}}\mathrm{e}^{-i(-\bm{k}'+\bm{q})\cdot\bm{n}} c_{-\bm{k}} c_{-\bm{k}'}^{\dag}
     \nonumber\\
          &=\sum_{\bm{k}}
\mathrm{e}^{-i(-\bm{k}+\bm{q})\cdot\bm{n}} c_{-\bm{k}} c_{-\bm{k}}^{\dag},
\end{align}
where $V$ is the volume of the system and $\bm{n}$ is an arbitrary vector in the real space.  
Therefore,  BdG Hamiltonian in the wave-vector space, when the superconductor is subjected to a current bias, has the structure:
\begin{align}
    \left[  
      \begin{array}{cc}
        {h}_{\bm{k}+\bm{q}} & \Delta  \\ 
        \Delta^\dag & -{h}^*_{-\bm{k}+\bm{q}}   
      \end{array}
     \right]\;.
\end{align}

\section*{Appendix C: Linear response of the Edelstein effect}
In this part, we derive the linear response function of the Edelstein effect as the responce function of the $q_x$ representing the supper current along $x$ direction.   
At first we change the basis by using 
\begin{align} \label{eq:ut1}
  U=\left[    
    \begin{array}{cc}
    \hat{L}_0 \tilde{\sigma}_0 & 0  \\ 
     0  &   \hat{L}_0 i\tilde{\sigma}_y 
   \end{array}
  \right],
\end{align}
where $\hat{L}_0$ is the unit matrix in the orbital space and $\tilde{\sigma}_y$ is the $y$ component of the Pauli matrix in the spin space.
Then the Hamiltonian is transformed as
  \begin{align}
  H'_{\bm{k},\bm{q}}&=
  U H_{\bm{k},\bm{q}} U^\dag\\
 &=\left[  
  \begin{array}{cc}
   H_N(\bm{k}+\bm{q}) & \mathcal{D}  (-i\tilde{\sigma}_y)\\ 
   \mathcal{D} i\tilde{\sigma}_y &  i\tilde{\sigma}_y (-H^*_N(-\bm{k}+\bm{q}))
   (-i\tilde{\sigma}_y)   
  \end{array}
 \right]\nonumber\\
 &=\left[  
  \begin{array}{cc}
   H_N(\bm{k}+\bm{q}) &  \Delta\hat{L}_0 \tilde{\sigma}_0 \\ 
   \Delta\hat{L}_0\tilde{\sigma}_0   &  -H_N(\bm{k}-\bm{q})   
  \end{array}
 \right],
\end{align}
where $\bm{q}=(q_x,0,0)$.
Here we use $\hat{\sigma}_y
\hat{\sigma}_{x,z}\hat{\sigma}_y=-\hat{\sigma}_{x,z}$ and 
$\hat{\sigma_y}^*=-\hat{\sigma_y}$ and $\hat{\bm{L}}^*=-\hat{\bm{L}}$.
Then we can expand  $H'_{\bm{k},\bm{q}}$ within the first order of $q_x$ as 
\begin{align}
  H'_{\bm{k},\bm{q}}= H^0_{\bm{k}}+H^1_{\bm{k}}q_x \label{eq:sr_hamiltonian}
\end{align}
where
\begin{align}
    H^0_{\bm{k}}&=\left[\begin{array}{cc}
   H_N(\bm{k}) &  \Delta\hat{L}_0 \tilde{s}_0 \\ 
   \Delta\hat{L}_0 \tilde{s}_0  &  -H_N(\bm{k})   
  \end{array}
 \right],\\
    H^1_{\bm{k}}&=\left[\begin{array}{cc}
   \partial_{k_x}H_N(\bm{k}) &  0 \\ 
   0  &  \partial_{k_x}H_N(\bm{k})   
  \end{array}
 \right].
\end{align}
When the Greens function expanded within the first order is given by 
\begin{align}
  {\mathcal{G}}_{i\omega}({\bm{k},\bm{q}})&=
    [i\omega-H_{\bm{k},\bm{q}}]^{-1}\nonumber\\
  &\simeq G(\bm{k},i\omega)+G(\bm{k},i\omega)
  H^1_{\bm{k}}G(\bm{k},i\omega)q_x,\\
  G(\bm{k},i\omega)&=[i\omega-H^0_{\bm{k}}]^{-1},
\end{align}
where $i\omega$ is Mastubara frequency.
When we define the orbital Edelstein susceptibility $\chi^o_{yx}$ as     
\begin{align}
  <L_y>  &= \chi^o_{yx}  q_x,\\  
  \chi^o_{yx}
  &=T\sum_{\bm{k},i\nu} \mathrm{Tr}
  [ v^s_x G(\bm{k},i\mathrm{\nu}) 
  L^s_y G(\bm{k},i\nu)],
\end{align}
where
\begin{align}
     v^s_x&=\partial_{k_x} \left[\begin{array}{cc}
   H_N(\bm{k}) &  0 \\ 
    0  &  -H_N(\bm{k})    \end{array} \right],\\
         L^s_x&=\partial_{k_x} \left[\begin{array}{cc}
   \hat{L}\tilde{\sigma}_0 &  0 \\ 
    0  &  \hat{L}\tilde{\sigma}_0   \end{array} \right].
\end{align}
Then we can expand $\chi^o_{yx}$ by Lehmann representation as following, 
\begin{align}
    \chi^o_{yx}&=T
    \sum_{\bm{k} i\nu}
    \mathrm{Tr}\left[  v^s_x G(\bm{k},i\nu) L^s_y G(\bm{k},i\nu)\right]\nonumber\\ 
    &=T
    \sum_\kappa 
    \bra{\kappa} \sum_{\bm{k}i\nu} v^s_x
    \sum_{\lambda p}
    \ket{\Phi_{\lambda\bm{k}}^p}
    \bra{\Phi_{\lambda\bm{k}}^p}
    [i\nu-E^{p}_{\lambda\bm{k}}]^{-1}\nonumber\\
    &\quad \times L^s_y  
    \sum_{\lambda'  p'}
    \ket{\Phi_{\lambda' \bm{k}}^{p'}}\bra{\Phi_{\lambda'\bm{k}}^{p'}}
    [i\nu-E^{p'}_{\lambda'\bm{k}}]^{-1} \ket{\kappa}\nonumber\\
   &= T \sum_{\bm{k} i\nu\lambda \lambda' p p' }  
    \bra{\Phi_{\lambda'\bm{k}}^{p'}}
    v^s_x
    \ket{\Phi_{\lambda \bm{k}}^p}\bra{\Phi_{\lambda \bm{k}}^p} L^s_y  
    \ket{\Phi_{\lambda' \bm{k}}^{p'}}\nonumber\\
     &\quad\times [i\nu-E^{p}_{\lambda\bm{k}}]^{-1}     
    [i\nu-E^{p}_{\lambda\bm{k}}]^{-1} \nonumber\\
    &=T \sum_{\bm{k} i\nu \lambda \lambda' p p' } 
    \bra{\phi_{\lambda \bm{k}}} v_x 
    \ket{ \phi_{\lambda'\bm{k}}}    
    |\braket{\Psi^{p}_{\lambda \bm{k}}| \Psi^{p'}_{\lambda'\bm{k}}}\nonumber\\
    &\quad\times\bra{\phi_{\lambda' \bm{k}}}L_y 
    \ket{ \phi_{\lambda\bm{k}}}    
    \braket{\Psi^{p'}_{\lambda' \bm{k}}| \Psi^{p}_{\lambda\bm{k}}}\nonumber\\
    &\quad \times  
    \frac{f(E^p_{\lambda \bm{k}})-f(E^{p'}_{\lambda'\bm{k}})}
       {E^p_{\lambda \bm{k} } -E^{p'}_{\lambda' \bm{k}}}  
    \left.\right]\nonumber \\   
    &=-2\sum_{k\lambda\lambda'}
\frac{\braket{\Psi^{h}_{\lambda \bm{k}} |\Psi^{e}_{{\lambda'} \bm{k}}}^2}{{E_{\lambda \bm{k}}+{E_{\lambda' \bm{k}}}}}
\bra{\phi_{\lambda \bm{k}}}
   v_x\ket{ \phi_{\lambda'\bm{k}}}
   \bra{\phi_{\lambda' \bm{k}}}
   L_y\ket{ \phi_{\lambda\bm{k}}}\nonumber\\
   &=-2\sum_{\bm{k}\lambda>\lambda'}
\frac{1}{{E_{\lambda \bm{k}}+{E_{\lambda'\bm{k}}}}}
\bra{\phi_{\lambda \bm{k}}}
   v_x\ket{ \phi_{\lambda'\bm{k}}}
   \bra{\phi_{\lambda' \bm{k}}}
   L_y\ket{ \phi_{\lambda\bm{k}}}\nonumber\\
   &\quad\times\left[  1-\frac{\varepsilon_{\lambda \bm{k}} \varepsilon_{\lambda' \bm{k}} +\Delta^2}{E_{\lambda \bm{k}}E_{\lambda' \bm{k}}}\right]  
   \label{eq:edelstein_calculation}
\end{align}
where
\begin{align}
  E^e_{\lambda \bm{k}}&=E_{\lambda \bm{k}}=\sqrt{\varepsilon^2_{\lambda \bm{k}}+\Delta^2}, \\
  E^h_{\lambda \bm{k}}&=-\sqrt{\varepsilon^2_{\lambda \bm{k}}+\Delta^2},
\end{align}
which is the eigenvalue of
\begin{align}
 & \ket{\Phi^e_{\lambda \bm{k}}}=
  \left[    
    \begin{array}{cc}
     u_{\lambda \bm{k}} \ket{\phi_{\lambda \bm{k}}}  \\ 
     v_{\lambda \bm{k}} \ket{\phi_{\lambda \bm{k}}}  
   \end{array}
  \right]=\ket{\Psi^e_{\lambda \bm{k}}}\otimes\ket{\phi_{\lambda \bm{k}}},\\
 & \ket{\Phi^h_{\lambda \bm{k}}}=
  \left[    
    \begin{array}{cc}
     -v_{\lambda \bm{k}} \ket{\phi_{\lambda \bm{k}}}  \\ 
     u_{\lambda \bm{k}} \ket{\phi_{\lambda \bm{k}}}  
   \end{array}
  \right]=  \ket{\Psi^h_{\lambda \bm{k}}}\otimes\ket{\phi_{\lambda \bm{k}}},
  \end{align}
  where
  \begin{align}
  &u_{\lambda \bm{k}}=\frac{1}{\sqrt{2}} \sqrt{1+\frac{\varepsilon_{\bm{k}\lambda}}{E_{\bm{k}\lambda}}} \label{eq:u},\\
  &v_{\lambda \bm{k}}=\frac{1}{\sqrt{2}} \sqrt{1-\frac{\varepsilon_{\bm{k}\lambda}}{E_{\bm{k}\lambda}}}, \label{eq:v}
\end{align}
and
\begin{align}
v_x &=\partial_{k_x} H_N(\bm{k}) ,\\
L_y&=\hat{L}_y\tilde{\sigma}_0.
\end{align}
Here $\ket{\phi_{\lambda \bm{k}}}$ is the eigenvector of the normal state $\varepsilon_{\bm{k}\lambda}$,  
$\ket{\Psi^{e(h)}_{\lambda \bm{k}}}$ is the eigenvector of the electron (hole) space, and  
$\ket{\kappa}$ contains spin, orbital and particle-hole spaces.  
In the calculation of Eq, (\ref{eq:edelstein_calculation}), we used
\begin{align}
     T\sum_{i\nu}[i\nu-E^{p}_{\lambda\bm{k}}]^{-1}     
    [i\nu-E^{p'}_{\lambda'\bm{k}}]^{-1}=
    \frac{f(E^p_{\lambda \bm{k}})-f(E^{p'}_{\lambda'\bm{k}})}
       {E^p_{\lambda \bm{k} } -E^{p'}_{\lambda' \bm{k}}}
\end{align}
where $f(E^e_{\lambda\bm{k}})-f(E^{h}_{\lambda'\bm{k}})=1$ 
and $f(E^{e(h)}_{\lambda\bm{k}})-f(E^{e(h)}_{\lambda'\bm{k}})=0$.
By replacing the $L_y$ to $s_y=\hat{L}_0\tilde{s}_y$ in Eq. (\ref{eq:edelstein_calculation}), we can get also the susceptibility of the spin Edelstein effect as
\begin{align}
\label{eqAppB:chiS}
    \chi^s_{xy}=&-2\sum_{\bm{k}\lambda>\lambda'}
\frac{1}{{E_{\lambda \bm{k}}+{E_{\lambda'\bm{k}}}}}
\bra{\phi_{\lambda \bm{k}}}
   v_x\ket{ \phi_{\lambda'\bm{k}}}
   \bra{\phi_{\lambda' \bm{k}}}
   s_y\ket{ \phi_{\lambda\bm{k}}}\nonumber\\
   &\quad\times\left[  1-\frac{\varepsilon_{\lambda \bm{k}} \varepsilon_{\lambda' \bm{k}} +\Delta^2}{E_{\lambda \bm{k}}E_{\lambda' \bm{k}}}\right].
\end{align}

\section*{Appendix D: Effective Hamiltonian to describe the Orbital Edestein effect with reduced twofold orbital sector}
In this part, we derive the effective 4x4 Hamiltonian,
which consists only of two bands with dominant $xy$ and $zx$ orbitals. In principle, the analysis can be performed with any pair of bands by projecting out the third one.
Here we define the diagonal component of the orbital basis 
Hamiltonian for $yz$, $zx$, and $xy$ component as $\varepsilon_{yz}$, $\varepsilon_{zx}$, and $\varepsilon_{xy}$, respectively. Then, we expect that the orbital Edelstein effect can be described by this model with good approximation 
if the difference of the energy between $yz(xy)$ orbital and $zx$ orbital is enough large compared with $\lambda$ and $\alpha$, i.e.
$|\varepsilon_{yz} (\varepsilon_{xy})-\varepsilon_{zx}|\gg |\alpha|,|\lambda|$. Indeed, the suppression of the orbital Edelstein effect due to inducing $\lambda$ and its sign change at the avoiding crossing are explained by this model well.   

The normal state Hamiltonian is written by
\begin{align}
  \mathcal{H}^{\mathrm{eff}}=\frac{1}{2}\sum_{\bm{k}}C^\dag_{\bm{k}} 
  H^{\mathrm{eff}}_N(\bm{k})C_{\bm{k}},
\end{align}
where
\begin{align}
&H^{\mathrm{eff}}_N(\bm{k})=\left[\begin{array}{cccc}
    \varepsilon_{yz} & -i\alpha(k_x) & 0 & -\lambda' \\ 
    i\alpha(k_x) & \varepsilon_{xy} & \lambda' & 0 \\  
    0 &  \lambda' &   \varepsilon_{yz} & -i\alpha(k_x) \\
    -\lambda' &  0 &    
    i\alpha(k_x) & \varepsilon_{xy} 
    \end{array} \right]\\
    &C^\dag_{\bm{k}}=[
  c^\dag_{\bm{k}\uparrow yz}
  c^\dag_{\bm{k}\uparrow xy}
  c^\dag_{\bm{k}\downarrow yz}
  c^\dag_{\bm{k}\downarrow xy}].
\end{align}
Here we use  $\alpha(k_x)=\alpha\sin(k_x)$ as   the orbital Rashba term and $\lambda'=\lambda/2$ as spin orbit coupling closed in the $yz$ and $xy$ orbital component.   

At first, we write eigenvectors of the 4x4 Hamiltonian as
\begin{align}
    \ket{u_1}=\frac{1}{\sqrt{4R_-(p+R_-)}}
    \left[\begin{array}{cccc}   
    p+R_-  \\ 
    i(\alpha(k_x)-\lambda') \\
    -i(p+R_-)\\
    (\alpha(k_x)-\lambda') \\
    \end{array} \right],\\
    \ket{u_2}=\frac{1}{\sqrt{4R_-(p+R_-)}}
    \left[\begin{array}{cccc}   
    i(\alpha(k_x)-\lambda')  \\ 
    p+R_- \\
    \alpha(k_x)-\lambda'\\
    -i(p+R_-) \\
    \end{array} \right],\\
    \ket{u_3}=\frac{1}{\sqrt{4R_+(p+R_+)}}
    \left[\begin{array}{cccc}   
    -i(p+R_+)  \\ 
    \alpha(k_x)+\lambda' \\
    p+R_+\\
    i(\alpha(k_x)+\lambda') \\
    \end{array} \right],\\
    \ket{u_4}=\frac{1}{\sqrt{4R_+(p+R_+)}}
    \left[\begin{array}{cccc}   
    \alpha(k_x)+\lambda'   \\ 
     -i(p+R_+)\\
    i(\alpha(k_x)+\lambda')\\
    p+R_+ \\
    \end{array} \right],
\end{align}
for the eigen values
\begin{align}
    & \varepsilon_1=q+R_-,\\
    & \varepsilon_2=q-R_-,\\
    & \varepsilon_3=q+R_+,\\
    & \varepsilon_4=q-R_+,
\end{align}
where 
\begin{align}
    q=&(\varepsilon_{yz}+\varepsilon_{xy})/2,\\
    p=&(\varepsilon_{yz}-\varepsilon_{xy})/2,\\
    R_{\pm}=&\sqrt{p^2+(\alpha(k_x)\pm \lambda')^2}.
\end{align}
Here the eigen vectors are not well defined in the case $p+R_\pm=0$. However we can avoid this problem by normalization after calculating the expectation value.  
    Then only the combination of $\ket{u_1}$,$\ket{u_2}$ and $\ket{u_3}$,$\ket{u_4}$ contribute
    orbital Edelstein effect (i.e. $\bra{u_1}L_y\ket{u_{3(4)}}=0$ and $\bra{u_2}L_y\ket{u_{3(4)}}=0$ ). 
    As a result,
\begin{align}
&-\bra{u_1} v_x \ket{ u_2}
   \bra{u_2} L_y \ket{ u_1}\nonumber\\
   &=\frac{p}{R_-^2}[(\alpha(k_x)-\lambda')\partial_{k_x}p-p\partial_{k_x}\alpha(k_x)]\nonumber\\
   &=\frac{p}{p^2+(\alpha(k_x)-\lambda')^2}[(\alpha(k_x)-\lambda')\partial_{k_x}p\nonumber\\ \quad&-p\partial_{k_x}\alpha(k_x)],\\
   &-\bra{u_3} v_x \ket{ u_4}
   \bra{u_4} L_y \ket{ u_3}\nonumber\\
   &=\frac{p}{R_+^2}[(\alpha(k_x)+\lambda')\partial_{k_x}p-p\partial_{k_x}\alpha(k_x)]\nonumber\\
   &=\frac{p}{p^2+(\alpha(k_x)+\lambda')^2}[(\alpha(k_x)+\lambda')\partial_{k_x}p\nonumber\\ \quad&-p\partial_{k_x}\alpha(k_x)].
\end{align}

When we focus on the vicinity of the avoiding crossing, namely considering the limit of $p=0$, the susceptibility of the orbital Edelstein effect is given by   
\begin{align}
   \chi^o_{xy}&= \frac{(\varepsilon_{yz}-\varepsilon_{xy})\partial_{k_x}(\varepsilon_{yz}-\varepsilon_{xy})}{4(\alpha(k_x)-\lambda')}\times A \nonumber\\
   &+\frac{(\varepsilon_{yz}-\varepsilon_{xy})\partial_{k_x}(\varepsilon_{yz}-\varepsilon_{xy})}{4(\alpha(k_x)+\lambda')}\times B    
   \label{eq:chi_first_order}
\end{align}
with
\begin{align}
    A&=\frac{1}{{E_{1 }+{E_{2}}}} \left[  1-\frac{\varepsilon_{1 } \varepsilon_{2 } +\Delta^2}{E_{1 } E_{2 }}\right],\\
   B&=\frac{1}{{E_{3 }+{E_{4}}}} \left[  1-\frac{\varepsilon_{3 } \varepsilon_{4 } +\Delta^2}{E_{3 } E_{4 }}\right],
\end{align}
where $E_i=\sqrt{\varepsilon_i^2+\Delta^2}$.

 Here A (B) is the function having finite value in the momentum space where $\varepsilon_1\times\varepsilon_2<0$ when $|\varepsilon_1-\varepsilon_2|\gg\Delta $. $\int A dk_x$ and $\int B dk_x$ are not so sensitive for $\lambda$. Therefore, we can discuss the properties of $\chi^o_{xy}$ by summation of  
\begin{align}
&\frac{(\varepsilon_{yz}-\varepsilon_{xy})(\partial_{k_x}(\varepsilon_{yz}-\varepsilon_{xy}))}{(\alpha(k_x)-\lambda')}
+\frac{(\varepsilon_{yz}-\varepsilon_{xy})(\partial_{k_x}(\varepsilon_{yz}-\varepsilon_{xy}))}{(\alpha(k_x)+\lambda')}\nonumber\\
  &=  \frac{2\alpha(k_x)}{\alpha^2(k_x)-\lambda'^2} (\varepsilon_{yz}-\varepsilon_{xy})(\partial_{k_x}(\varepsilon_{yz}-\varepsilon_{xy})).
\end{align}
Then we can see that the orbital Edelstein effect is zero at $\varepsilon_{yz}-\varepsilon_{xy}=0$ region on the Fermi energy and has different sign between $\varepsilon_{yz}-\varepsilon_{xy}>0$ region and $\varepsilon_{yz}-\varepsilon_{xy}<0$ region at the Fermi energy. 
Furthermore, larger $|\lambda'|$ more increase $|\chi^o_{xy}|$ in  
$|\alpha(k_x)|<|\lambda'|$ region and
when $|\lambda'|$ terms larger $|\alpha(k_x)|$, $\chi^o_{xy}$ has sign change.

Next we extract $\lambda$ dependent parts from 
$-\bra{u_1} v_x \ket{ u_2}
   \bra{u_2} L_y \ket{ u_1}$ and $-\bra{u_3} v_x \ket{ u_4}
   \bra{u_4} L_y \ket{ u_3}$ to discuss the $\lambda'$ dependence of $\chi^o_{xy}$, which are  
\begin{align}
&\frac{p}{p^2+(\alpha(k_x)-\lambda')^2}(\alpha(k_x)-\lambda')\partial_{k_x}(\varepsilon_{yz}-\varepsilon_{xy})/2\label{eq:first_term_1}\\
  & \frac{p}{p^2+(\alpha(k_x)+\lambda')^2}(\alpha(k_x)+\lambda')\partial_{k_x}(\varepsilon_{yz}-\varepsilon_{xy})/2\label{eq:first_term_2}.
\end{align}
In $|p|>|\alpha(k_x)\pm \lambda' |$ region, 
we can expand above terms as 
\begin{align}
    &\frac{p(\alpha(k_x)\pm \lambda')}{p^2+(\alpha(k_x)\pm \lambda')^2}\nonumber\\
    &=\frac{\alpha(k_x)\pm\lambda' }{p}- \frac{(\alpha(k_x)\pm\lambda')^3 }{p^3} +O(\alpha(k_x)\pm \lambda')^4)
\end{align}
within the third order of $\alpha(k_x)\pm\lambda'$.   
Therefore, we can see one contribute as the suppression of $\chi^o_{xy}$ and another contribute increasing of  $\chi^o_{xy}$ in Eqs (\ref{eq:first_term_1})(\ref{eq:first_term_2}). However we can see the suppression effect is larger comparing the third order terms. 

For instance, focusing on $\alpha(k_x)>\lambda'>0$ region, we can see 
Eq. (\ref{eq:first_term_1}) term contributes to suppress the absolute value of the Edelstein effect, on the other hand, Eq. (\ref{eq:first_term_2}) term contributes to increase the absolute value of the Edelstein effect within the first order of $\alpha(k_x)\pm \lambda'$. Including the third order, we can see the suppression effect of Eq. (\ref{eq:first_term_1}) is larger than
incleasing effect of Eq. (\ref{eq:first_term_2}) because the $
-(\alpha(k_x)-\lambda')^3-(\alpha(k_x)+\lambda')^3=-(2\alpha(k_x)^3+6\alpha(k_x)\lambda'^2)$,
which contribute to reduce $|\chi^o_{xy}|$.

\section*{Appendix E: Downfolded Hamiltonian to describe the Spin Edelstein effect }
In this section, we derive a downfolded 2x2 Hamiltonian  
for any given orbital dependent band to discuss the emergent spin Edelstein effect. To thus aim we employ the L\"owdin downfolding technique. 
At first, we consider the Hamiltonian by focusing on the $yz$ orbital state including the effect of the $zx$ and $xy$ orbital as orbitally hybridized states through the spin-orbit and orbital Rashba interaction.
Then, one can rewrite the normal Hamiltonian, 
\begin{align}
    H_N=C^\dag_{\bm{k}} H_N(\bm{k})  C_{\bm{k}}
\end{align}
where
\begin{align}
&H_N(\bm{k})=
    \left[\begin{array}{cccccc} 
    \varepsilon_{yz} & 0 & -i \lambda' & 0 & -i\alpha_x& -\lambda'   \\
     0 &\varepsilon_{yz} & 0 & i\lambda' &\lambda' &-i\alpha_x \\
     i \lambda'  & 0 &\varepsilon_{zx} &0 & -i \alpha_y &i\lambda'\\
     0 & -i \lambda' & 0 & \varepsilon_{zx} &i\lambda' &-i\alpha_y\\
     i\alpha_x&\lambda'&i \alpha_y&-i\lambda'&\varepsilon_{xy}&0\\
     -\lambda'&i\alpha_x&-i\lambda'&i \alpha_y&0&\varepsilon_{xy}
    \end{array} \right],\nonumber\\ 
   & C^\dag_{\bm{k}}=[
  c^\dag_{\bm{k}\uparrow yz}
  c^\dag_{\bm{k}\downarrow yz}
  c^\dag_{\bm{k}\uparrow zx}
  c^\dag_{\bm{k}\downarrow zx}
    c^\dag_{\bm{k}\uparrow xy}
  c^\dag_{\bm{k}\downarrow xy}].
\end{align}
Here we define $\alpha_x=\alpha\sin(k_x)$ and $\alpha_y=\alpha\sin(k_y)$. $\lambda'=\lambda/2$. 
If we define 
\begin{align}
    V=
    \left[\begin{array}{cccc} 
     -i \lambda' & 0 & -i\alpha_x& -\lambda'   \\
      0 & i\lambda' &\lambda' &-i\alpha_x 
    \end{array} \right]
\end{align}
and 
\begin{align}
    H_B=
    \left[\begin{array}{cccc} 
     \varepsilon_{zx} &0 & 0&i\lambda'\\
     0 & \varepsilon_{zx} &i\lambda' &0\\
     0&-i\lambda'&\varepsilon_{xy}&0\\
     -i\lambda'&0&0&\varepsilon_{xy}
    \end{array} \right],
\end{align}
the 2x2 effective Hamiltonian summarized into $yz$ basis is given by the following equation where we have been using the expansion at the second order of $V$,
\begin{align}
    H_{2\times2}=
\left[\begin{array}{cc} 
     \varepsilon_{yz} &0 \\
     0 & \varepsilon_{yz} \\
    \end{array} \right]+
    V(\varepsilon-H_B)^{-1}V^\dag,
\end{align}
where $\varepsilon$ is a eigenvalue of the $H_N(\bm{k})$. Now we do not mention the detail of $\varepsilon$ because here we seek the qualitative description of spin Edelstein effect. Then 2x2 Hamiltonian becomes
\begin{align}
     H_{2\times2}&=
     h_0s_0 +h_x s_x +h_y s_y +h_z s_z
     \label{eq:two_by_two}
\end{align}
\begin{align}
&h_0=\varepsilon_{yz} +\frac{\alpha_x^2 \varepsilon'_{zx}[-\alpha_y^2+\varepsilon'_{zx}\varepsilon'_{xy}-\lambda'^2]}{\Theta}\nonumber\\
&+\frac{\lambda'^2[-\alpha_y^2(\varepsilon'_{yz}+\varepsilon'_{xy}+2\lambda')]
}{\Theta}\nonumber\\
&+\frac{\lambda'^2(\varepsilon'_{xy}+\varepsilon'_{zx}-2\lambda')(\varepsilon'_{xy}\varepsilon'_{zx}-\lambda'^2)}{\Theta},\\
&h_x=\frac{2\lambda'\alpha_y(-\alpha_x^2\varepsilon'_{zx}+\lambda' (\alpha_y^2+(\varepsilon'_{xy}-\lambda')(-\varepsilon'_{zx}+\lambda'))
)}{\Theta}\\
&h_y=\frac{-2\lambda' \alpha_x[\alpha_y^2(\varepsilon'_{zx}+\lambda')-(\varepsilon'_{zx}-\lambda')(\varepsilon'_{xy}\varepsilon'_{zx}-\lambda'^2)]}{\Theta}\\
&h_z=0
\end{align}
where 
\begin{align}
\varepsilon'_{xy(zx)}=&(\varepsilon-\varepsilon_{xy(zx)}),\\
\Theta=&\alpha_y^4+(-\varepsilon'_{xy}\varepsilon'_{zx}+\lambda'^2)^2-2\alpha_y^2(\varepsilon'_{xy}\varepsilon'_{yz}+\lambda'^2).
\end{align}
Next we show the 2x2 Hamiltonian with the $zx$ basis.
\begin{align}
&h_0=\varepsilon_{zx} +\frac{\alpha_y^2 \varepsilon'_{yz}[-\alpha_x^2+\varepsilon'_{xy}\varepsilon'_{yz}-\lambda'^2]}{\Theta}\nonumber\\
&+\frac{\lambda'^2[-\alpha_x^2(\varepsilon'_{xy}+\varepsilon'_{zx}-2\lambda')]}{\Theta}\nonumber\\
&+\frac{\lambda'^2(\varepsilon'_{xy}+\varepsilon'_{yz}+2\lambda')(\varepsilon'_{xy}\varepsilon'_{yz}-\lambda'^2)}{\Theta},\\
&h_x=\frac{2\lambda' \alpha_y[\alpha_x^2(\varepsilon'_{yz}-\lambda')-(\varepsilon'_{yz}+\lambda')(\varepsilon'_{xy}\varepsilon'_{yz}-\lambda'^2)]}{\Theta}\\
&h_y=\frac{-2\lambda'\alpha_x(\alpha_y^2\varepsilon'_{yz}+\lambda' (\alpha_y^2-(\varepsilon'_{xy}+\lambda')(\varepsilon'_{yz}+\lambda'))
)}{\Theta}\\
&h_z=0
\end{align}
where 
\begin{align}
\varepsilon'_{xy(yz)}=&(\varepsilon-\varepsilon_{xy(yz)}),\\
\Theta=&\alpha_x^4+(-\varepsilon'_{xy}\varepsilon'_{yz}+\lambda'^2)^2-2\alpha_x^2(\varepsilon'_{xy}\varepsilon'_{yz}+\lambda'^2).
\end{align}
At last we show the 2x2 Hamiltonian with the $xy$ basis.
\begin{align}
    &h_0=\varepsilon_{xy} +\frac{\alpha_y^2\varepsilon'_{yz}+\alpha_x^2\varepsilon'_{zx}+(\varepsilon'_{yz}+\varepsilon'_{zx}-2\lambda')\lambda'^2}{\varepsilon'_{yz}\varepsilon'_{zx}-\lambda'^2}    
    \\
&h_x=\frac{-2\lambda'\alpha_y(\varepsilon'_{yz}-\lambda')}{\varepsilon'_{yz}\varepsilon'_{zx}-\lambda'^2}\\
&h_y=\frac{2\lambda'\alpha_x(\varepsilon'_{zx}-\lambda')}{\varepsilon'_{yz}\varepsilon'_{zx}-\lambda'^2}\\
&h_z=0.
\end{align}
For all the derived 2x2 Hamiltonians,
$h_0$ is even function for $\bm{k}$ and $h_z$=0.
Adiitionally, $h_x$ is $k_y$ odd and $k_x$ even and $h_y$ is $k_x$ odd and $k_y$ even due to the time reversal symmetry.
When we derive the spin susceptibility,
\begin{align}
    \bra{u_+}\partial_{k_x} H_{2\times 2}\ket{u_-}
\end{align}
and
\begin{align}
\bra{u_-} s_y\ket{u_+}    
\end{align}
with
\begin{align}
    \ket{u_+}=&\frac{1}{\sqrt{2}}
    \left[\begin{array}{cc} 
     1 \\
      (h_x+ih_y)/h
    \end{array} \right],\\
    \ket{u_-}=&\frac{1}{\sqrt{2}}
    \left[\begin{array}{cc} 
     1 \\
      -(h_x+ih_y)/h
    \end{array} \right],
\end{align}
where $h=\sqrt{h_x^2+h_y^2}$ and $\ket{u_\pm}$ is the eigenvector of $H_{2\times2}$,
\begin{align}
    \bra{u_-} s_y\ket{u_+} = -ih_x/h
\end{align}
and
\begin{align}
    \bra{u_+}\partial_{k_x} H_{2\times 2} \ket{u_-} = \frac{i}{h}[h_x\partial_{k_x}h_y-h_y\partial_{k_x}h_x]. 
\end{align}
Therefore,  
\begin{align}
    \chi^s_{yx}=&\frac{2}{h^2}[-h_x^2\partial_{k_x}h_y+h_xh_y\partial_{k_x}h_x]\nonumber\\
    &\times\frac{1}{{E_{+ }+{E_{-}}}} \left[  1-\frac{\varepsilon_{+ } \varepsilon_{- } +\Delta^2}{E_{+ } E_{- }}\right].
\end{align}
where
\begin{align}
\varepsilon_{\pm}=h_0\pm\sqrt{h^2_x+h^2_y},\\
E_{\pm}=\sqrt{\varepsilon^2_\pm+\Delta^2}.
\end{align}
By substitute the each 2x2 Hamiltonian, 
we can expect the spin Edelstein effect becomes finite value except $k_y=-\pi,0,\pi$, where spin Edelstein Effect is odd function for $\alpha$. Especially, around the gamma point for the band bottom originated from $xy$ orbital, we can expect spin Edelstein susceptibility is $ \lambda$ odd function when $\varepsilon'_{yz}$ and $\varepsilon'_{zx}$ is enough larger than $\lambda$.

\section*{Appendix F: Downfolded Hamiltonian with two orbital degrees of freedom to describe the spin Edelstein effect}

In this appendix we use the L\"owdin downfolding method to derive an effective 
Hamiltonian with four configurations, which includes the effect of the  hybridization between the ${yz}$ and ${xz}$ orbitals.
Within this description, we can show that a non zero spin Edelstein effect can be obtained at $k_y=0,\pm\pi$ and we can track the evolution of the spin Edelstein moment for electron fillings corresponding to the $xz,yz$ bands being occupied.
Here we set $\alpha_y=0$, $\alpha_x=\alpha$ and $\lambda'=\lambda/2$ for simplicity. 
The effective 4x4 Hamiltonian is then given by
\begin{align}
    H_{4\times4}=
    \left[\begin{array}{cccc} 
     \epsilon_{yz} &-2i\gamma & -i \eta&-\gamma\\
     2i \gamma&\epsilon_{yz}&-\gamma&i \eta\\
     i\eta &-\gamma& \epsilon_{zx}&0\\
     -\gamma& -i \eta& 0&\epsilon_{zx}
    \end{array} \right]
\end{align}
where
\begin{align}
\epsilon_{yz}&=\varepsilon_{yz}+\varepsilon'^{-1}_{xy}(\alpha^2+\lambda'^2)\\
\epsilon_{zx}&=\varepsilon_{zx}+\varepsilon'^{-1}_{xy}\lambda'^2\\
\eta&=\lambda'-\varepsilon'^{-1}_{xy}\lambda'^2\\
\gamma&=\alpha \lambda'\varepsilon'^{-1}_{xy}\\
\varepsilon'_{xy}&=\varepsilon-\varepsilon_{xy}
\end{align}
This Hamiltonian can be analitically  diagonalized, giving the following solutions for the eigenvalues 
\begin{equation}
    \varepsilon_{1,4}=q'\pm R',\quad  \varepsilon_{2,3}=q\pm R,\quad 
\end{equation}and 
eigenvectors
\begin{align}
   \ket{u_1}=
   \left[\begin{array}{cccc} 
     (p'+R')/r'\\
     i(p+R)/r\\
     ia/r\\
     a'/r'
    \end{array} \right]\\
    \ket{u_2}=
   \left[\begin{array}{cccc} 
     i(p'+R')/r'\\
     (p+R)/r\\
     a/r\\
     ia'/r'
    \end{array} \right]\\
    \ket{u_3}=
   \left[\begin{array}{cccc} 
     -ia'/r'\\
     -a/r\\
     (p+R)/r\\
     i(p'+R')/r'
    \end{array} \right]\\
     \ket{u_4}=
   \left[\begin{array}{cccc} 
     -a'/r'\\
     -ia/r\\
     i(p+R)/r\\
     (p'+R')/r'
    \end{array} \right]\\
\end{align}
where
\begin{align}
    a&=-\gamma+\eta\\
    q&=(\epsilon_{yz}+\epsilon_{zx}+2\gamma)/2\\
    p&=(\epsilon_{yz}-\epsilon_{zx}+2\gamma)/2\\
    R&=\sqrt{a^2+p^2}\\
    r&=\sqrt{4R(p+R)}\\
    a'&=-\gamma-\eta\\
    q'&=(\epsilon_{yz}+\epsilon_{zx}-2\gamma)/2\\
    p'&=(\epsilon_{yz}-\epsilon_{zx}-2\gamma)/2\\
    R'&=\sqrt{a'^2+p'^2}\\
    r'&=\sqrt{4R'(p'+R')}.
\end{align}
To analyze the spin Edelstein effects we need the matrix elements of $s_y$ and the velocity $\partial_{k_x}H_{4\times 4}$:
\begin{eqnarray}
    \bra{u_4}s_y\ket{u_1}&=&-[a'(p+R)+a(p'+R')]/rr',\\
    \bra{u_3}s_y\ket{u_2}&=&[a'(p+R)+a(p'+R')]/rr'
\end{eqnarray}
and
\begin{align}
        &\bra{u_1}\partial_{k_x}{H_{4\times4}}\ket{u_4}\nonumber\\=
    &[(a^2 +a'^2) \partial_{k_x}\gamma \nonumber\\
    &-( a (p + R)+a'(p'+R'))\partial_{k_x}(\epsilon_{yz} - \epsilon_{zx})\nonumber\\
    &- ((p + R)^2+(p'+R')^2
    ) \partial_{k_x}\gamma \nonumber\\
   &+ 2(-a a'+(p+R)(p'+R'))\partial_{k_x}\eta\nonumber\\
   &-2(a'(p+R)+a(p'+R'))\partial_{k_x}\gamma) ]/rr'\\
   \nonumber\\
          &\bra{u_2}\partial_{k_x}{H_{4\times4}}\ket{u_3}\nonumber\\=
          &[(a^2 +a'^2) \partial_{k_x}\gamma \nonumber\\
            &-( a (p + R)+a'(p'+R'))\partial_{k_x}(\epsilon_{yz} - \epsilon_{zx})\nonumber\\
            &- ((p + R)^2+(p'+R')^2
    ) \partial_{k_x}\gamma \nonumber\\
    &- 2(-a a'+(p+R)(p'+R'))\partial_{k_x}\eta\nonumber\\
    &+2(a'(p+R)+a(p'+R'))\partial_{k_x}\gamma) ]/rr'
\end{align}
Indeed, looking at the expression for $\chi^s_{xy}$ (see \eqref{eqAppB:chiS}), we have to evalutate terms like $c=\bra{u_1}\partial_{k_x}{H_{4\times4}}\ket{u_4}  \bra{u_4}s_y\ket{u_1}$. It is useful to perform an expansion to leading order in $\lambda$, to capture the essential features
\begin{eqnarray}
 c &=& c_0+c_1 \lambda+c_2\lambda^2 +c_3\lambda^3+O(\lambda^4)\\
 c_0&=&0,\qquad 
  c_1=0\\
  c_2&=&\! \frac{\alpha^2 }{4\varepsilon'_{xy}}\left[(\alpha^2\varepsilon'^{-1}_{xy}-\delta_\varepsilon +|\delta_\varepsilon+\alpha^2 \varepsilon'^{-1}_{xy}|)\right.\nonumber\\
 &\times&\left.\partial_{k_x}\varepsilon'^{-1}_{xy}+2 \varepsilon'^{-1}_{xy} \partial_{k_x}\delta_\varepsilon \right](\delta_\varepsilon+\alpha^2 \varepsilon'^{-1}_{xy})^{-2}\\
   c_3&=&-\frac{\alpha\varepsilon'^{-1}_{xy}\partial_{k_x}\varepsilon'^{-1}_{xy}}{4(\delta_\varepsilon+\alpha^2 \varepsilon'^{-1}_{xy})^3}\left\{-3\alpha^4 \varepsilon'^{-2}_{xy}\right.\nonumber\\
   &+&\left. \alpha^2 \varepsilon'^{-1}_{xy}[2(\alpha-\delta_\varepsilon)+|\delta_\varepsilon+\alpha^2 \varepsilon'^{-1}_{xy}|  ]\right.\nonumber\\
   &+&\left.(2\alpha+\delta_\varepsilon)(\delta_\varepsilon+ |\delta_\varepsilon+\alpha^2 \varepsilon'^{-1}_{xy}|)\right\}
\end{eqnarray}
where $\delta_\varepsilon=\varepsilon_{yz}-\varepsilon_{zx}$. Without entering fully in the details of the coefficients, it is crucial that the first term of the expansion is $O(\lambda^2)$.


\section*{Appendix G: Spin and orbital Edelstein effect at $k_y/\pi=0.7$}
\begin{figure}[htbp]
\label{fig:ap1}
\includegraphics[width=1\linewidth]{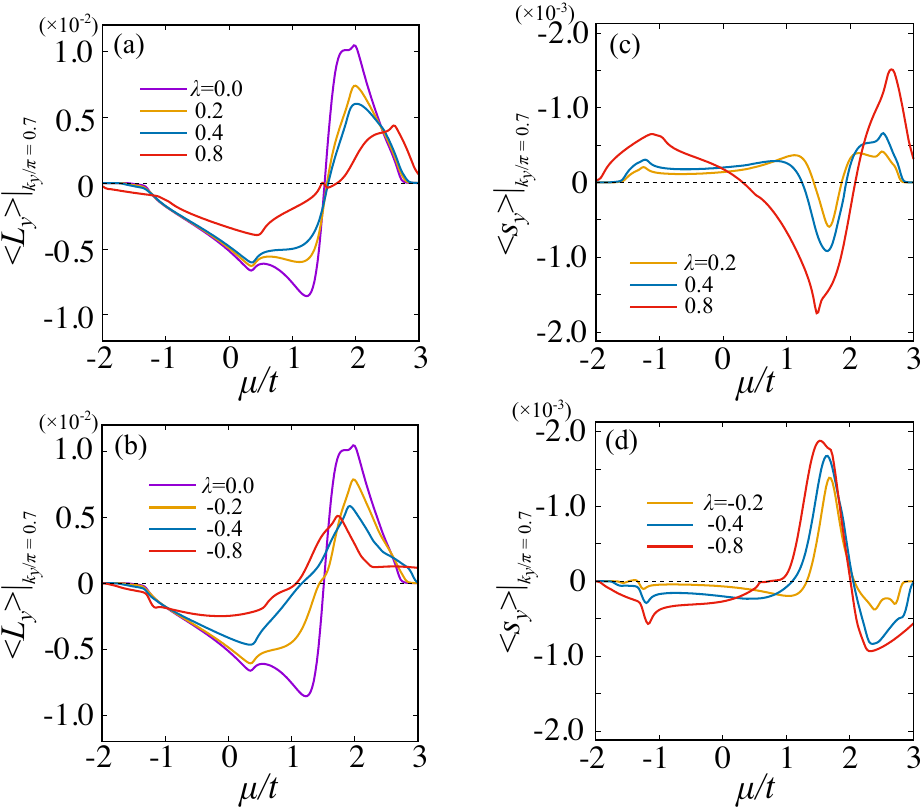}
\caption{Energy dependent profile, through the scan in $\mu$, of the induced orbital Edelstein moment $L_y$ for several values of the spin-orbit coupling  at $k_y/\pi=0.7$ line: (a) $\lambda \geq 0$ and  (b) $\lambda \leq 0$ in the superconducting state.
Energy dependent profile, through the scan in $\mu$, of the induced orbital Edelstein moment $s_y$ in the superconducting state for several values of the spin-orbit coupling  at $k_y/\pi=0.7$ line: (c) $\lambda \geq 0$ and  (d) $\lambda \leq 0$. }
\label{fig:figure6}
\end{figure}
In this appendix we show the energy dependent profile of the orbital (in Fig. 6 (a) (b)) and spin (in Fig. 6 (c) (d)) Edelstein effect at $k_y/\pi=0.7$ for $\lambda>0$ and $\lambda<0$.
We can see that a sign change in energy occurs for the orbital Edelstein effect as for smaller $k_y$ configurations. The spin contribution can be lager due to lager $k_y$ and the occurrence of multiple bands.


\begin{figure*}[htbp] 
\includegraphics[width=1\linewidth]{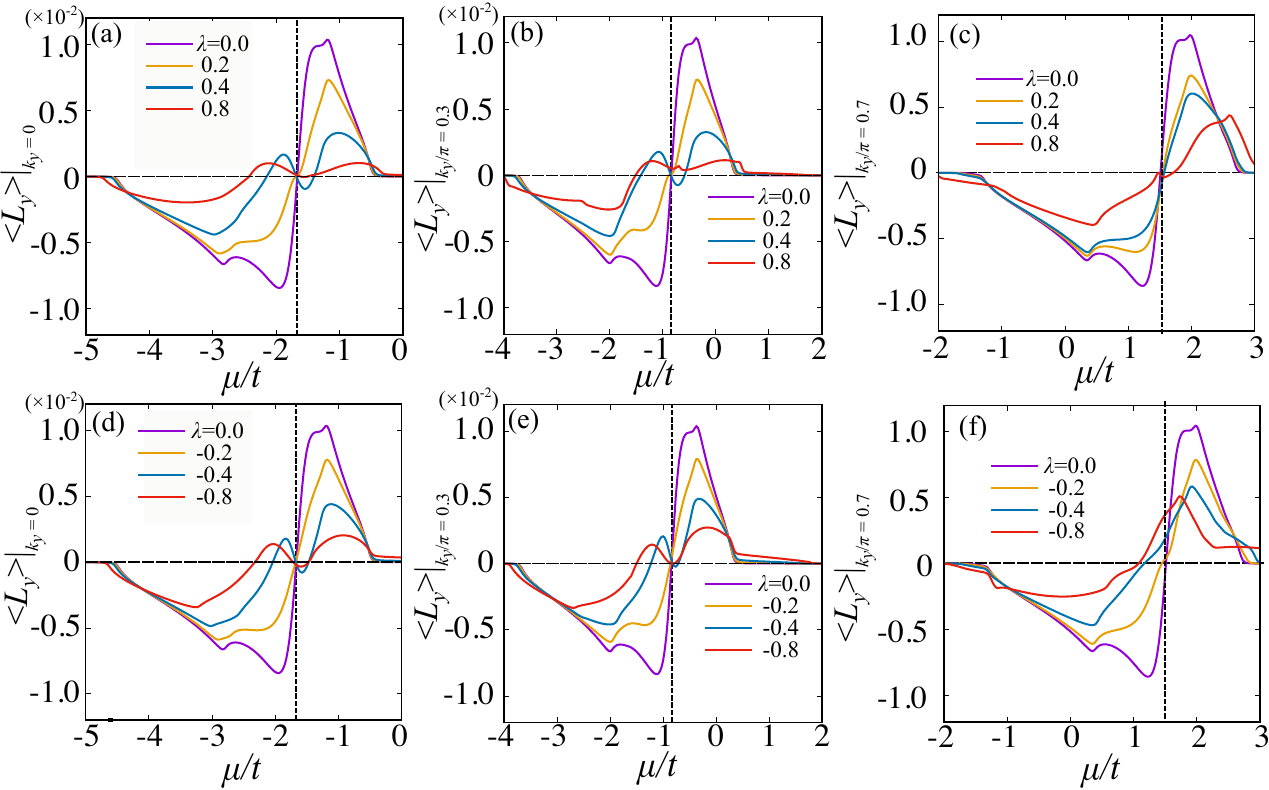}
\caption{
Energy dependent profile, through the scan in $\mu$, of the induced orbital Edelstein moment $L_y$ for several values of the spin-orbit coupling  and $k_y$ lines: (a) for $k_y=0$ (b) for $k_y/\pi=0.3$ (c) for $k_y/\pi=0.7$ with $\lambda \geq 0$.
(d) for $k_y=0$ (e) for $k_y/\pi=0.3$ (f) for $k_y/\pi=0.7$ with $\lambda \leq 0$. The black dotted lines show the energy positions of the crossing points between $\varepsilon_{yz}$ and $\varepsilon_{xy}$ for $\lambda=0$ and $\alpha=0$.  
} 
\label{fig:figure8}
\end{figure*}
\clearpage
In this appendix we show the energy dependent profile for the orbital Edelstein effect for $\lambda>0$ in Fig. \ref{fig:figure8}(a)(b)(c) and $\lambda<0$ in Fig. \ref{fig:figure8}(d)(e)(f) on several $ky$ lines. We see the energy profiles shift keeping those outlines with the band crossing point shifts due to $k_y$ change.



%


\end{document}